\documentclass[preprint,showpacs,showkeys,aps]{revtex4}
\usepackage{graphicx}% Include figure files
\usepackage{dcolumn}% Align table columns on decimal point
\usepackage{bm}% bold math
\begin{document}
\newcommand{\ti}[1]{\mbox{\tiny{#1}}}
\newcommand{\im}{\mathop{\mathrm{Im}}}
\def\be{\begin{equation}}
\def\ee{\end{equation}}
\def\bea{\begin{eqnarray}}
\def\eea{\end{eqnarray}}
\newcommand{\il}{~}
\title{Circular motion of neutral test particles in Reissner-Nordstr\"om spacetime}

\author{Daniela Pugliese}
\email{daniela.pugliese@icra.it}
\affiliation{Dipartimento di Fisica, Universit\`a di Roma "La Sapienza", Piazzale Aldo Moro 5, I-00185 Roma, Italy\\
ICRANet, Piazzale della Repubblica 10, I-65122 Pescara, Italy.}

\author{Hernando Quevedo}\thanks{On sabbatical leave from Instituto de Ciencias Nucleares, Universidad Nacional Aut\'onoma de M\'exico}
\email{quevedo@nucleares.unam.mx}
\affiliation{Dipartimento di Fisica, Universit\`a di Roma "La Sapienza", Piazzale Aldo Moro 5, I-00185 Roma, Italy\\
ICRANet, Piazzale della Repubblica 10, I-65122 Pescara, Italy.}

\author{Remo Ruffini}
\email{ruffini@icra.it}
\affiliation{Dipartimento di Fisica, Universit\`a di Roma "La Sapienza", Piazzale Aldo Moro 5, I-00185 Roma, Italy\\
ICRANet, Piazzale della Repubblica 10, I-65122 Pescara, Italy.}

\date{\today}

%\address{*ICRA—International Center for Relativistic Astrophysics, Physics
%Department (G9), University of Rome, ``La Sapienza'', Piazzale Aldo
%Moro 5, 00185 Rome, Italy  **ICRANET - C.C. Pescara, Piazzale della
%Repubblica 10, I-65100, Pescara, Italy,
%  }\email{*@, **@, ***@,****@}
\begin{abstract}
We investigate the motion of neutral test particles in the gravitational field of
a mass $M$ with charge $Q$ described by the Reissner-Nordstr\"om (RN) spacetime. We focus on
the study of circular stable and unstable orbits around configurations describing
either black holes or naked singularities. We show that at the classical radius, defined as $Q^2/M$,  there
exist orbits with zero angular momentum due to the presence of repulsive gravity.
The analysis of the stability of circular orbits indicates that black holes are characterized
by a continuous region of stability. In the case of naked singularities, the region of
stability can split into two non-connected regions inside which test particles move along
stable circular orbits.

\end{abstract} 
%97.10.Wn	Proper motions and radial velocities (line-of-sight velocities); space motions (see also 95.10.Jk Astrometry and reference systems)
%97.10.Gz	Accretion and accretion disks

%%% ----------------------------------------------------------------------
\maketitle
%%% ----------------------------------------------------------------------

%%%%%%%%%%%%%%%%%%%%%%%%%%%%%%%%%%%%%%%%%%%%%%%%%%%%%%%%%%%%
\section{Introduction}
In general relativity, the gravitational field of a static, spherically symmetric, charged body with mass $M$ and charge
$Q$ is described by the Reissner-Nordstr\"om  (RN) metric which in standard spherical coordinates can be expressed as
\begin{equation}\label{11metrica}
ds^2=-\frac{\Delta}{r^2}dt^2+\frac{r^2}{\Delta}dr^2
+r^2\left(d\theta^2+\sin^2\theta d\phi^2\right)\ ,
\end{equation}
where $\Delta=r^2-2Mr +Q^2$, and the associated electromagnetic
potential and field are
\begin{equation}\label{EMF}
A=\frac{Q}{r}dt,\quad F=dA=-\frac{Q}{r^2}dt\wedge dr \ ,
\end{equation}
respectively. The horizons are situated at $r_{\pm}=M\pm\sqrt{M^2-Q^2}$.

The study of the motion of test particles in this gravitational field is simplified 
by the fact that any plane through the center of the spherically symmetric gravitational 
source is a geodesic plane. Indeed, it can easily be seen that if the initial position and the tangent
vector of a geodesic lie on a plane that contains the center of the body, then the entire geodesic must 
lie on this plane. Without loss of generality we may therefore restrict ourselves to the study of
equatorial geodesics with $\theta =\pi/2$.

The tangent   vector $u^{a}$ to a curve $x^\alpha(\tau)$ is
$
u^{\alpha}={dx^{\alpha}}/{d\tau}=\dot{x}^{\alpha}$,
where $\tau$ is an affine parameter along the curve.
The momentum $p^\alpha= \mu\dot x^\alpha$ of a particle with  mass $\mu$
can be normalized so that
$g_{\alpha\beta}\dot{x}^{\alpha}\dot{x}^{\beta}=-k$, where $k=0,1,-1$ for null, timelike,
and spacelike curves, respectively. For the RN metric we obtain
\begin{equation}\label{esplinormacond}
-\frac{\Delta}{r^2}\dot{t}^{2}+\frac{r^2}{\Delta}\dot{r}^2
+r^2\dot{\phi}^{2}=-k \
\end{equation}
on the equatorial plane. The last equation reduces to a first-order differential equation
\begin{equation}\label{esplinormacondEL}
-\frac{E^2r^2}{\mu^{2}\Delta}+\frac{r^2 }{\Delta}\dot{r}^2+\frac{L^2}{\mu^{2}r^2}=-k  \ ,
\end{equation}
where we have used the expressions for the energy,  $E \equiv
-g_{\alpha\beta}\xi_{t}^{\alpha} p^{\beta}
=\mu\frac{\Delta}{r^2}\dot{t}$, and angular momentum,
$ L \equiv
g_{\alpha\beta}\xi_{\phi}^{\alpha}p^{\beta} = \mu r^2\dot{\phi}$ of
the test particle which are constants of motion associated with the
Killing vector fields $\xi_{t}=\partial_{t} $ and $\xi_{\phi}=\partial_{\phi} $,
respectively. Equation (\ref{esplinormacondEL}) can be rewritten as
\be
\dot r^2 + V^2 = \frac{E^2}{\mu^2}\ , \quad {\rm with} \quad
V\equiv\sqrt{\left(k+\frac{L^2}{\mu^2r^2}\right)\left(1-\frac{2M}{r}+\frac{Q^2}{r^2}\right)} \ .
\label{peff}
\ee
The investigation of the motion of test particles in the gravitational field of the RN metric is
thus reduced to the study of motion in the effective potential $V$. In this work, we will focus on the study of circular
orbits for which $\dot r = 0$ and $V=E/\mu$, with the condition $\partial V/\partial r =0$. A straightforward calculation
shows that this condition leads to
\be
\frac{L^2}{\mu^2} = k\frac{r^2(Mr-  Q^2)}{r^2 - 3Mr + 2Q^2} \ ,
\label{23}
\ee
an expression which we substitute in Eq.\il(\ref{peff}) to obtain
(an alternative analysis using an orthonormal frame is presented in Appendix\il\ref{AppA})
\be
\label{df}
\frac{E^2}{\mu^2} = k \frac{(r^2-2Mr+Q^2)^2}{r^2(r^2-3M r + 2Q^2)} \ .
\ee
Moreover, from the physical viewpoint it is important to find the minimum radius for stable circular orbits which is determined
by the inflection points  of the effective potential function, i.e., by  the condition $ {\partial^2 V}/{\partial^2 r}=0$. It is easy
to show that for the potential (\ref{peff}), the last condition is equivalent to
\begin{equation}\label{fgh}
M r^3-6 M^2 r^2+9 M Q^2 r-4 Q^4=0 \ .
\end{equation}

In this work, we present a detailed analysis of the circular motion of test particles governed by the above equations.
We will see that the behavior of test particles strongly depends on the ratio $Q/M$ and, therefore, we consider separately the case of
black holes, extreme black holes and naked singularities.

%%%%%%%%%%%%%%%%%%%%%%%%%%%%%%%%%%%%%%%%%%%%%%%%%%%%%%%%%%%%%%%%%%%%%%%%%%
%%%%%%%%%%%%%%%%%%%%%%%%%%%%%%%%%%%%%%%%%%%%%%%%%%%%%%%%%%%%%%%%%%%%%%%%%
\section{Black holes}

\begin{figure}%[h!]
\centering
\begin{tabular}{c}
\includegraphics[scale=1]{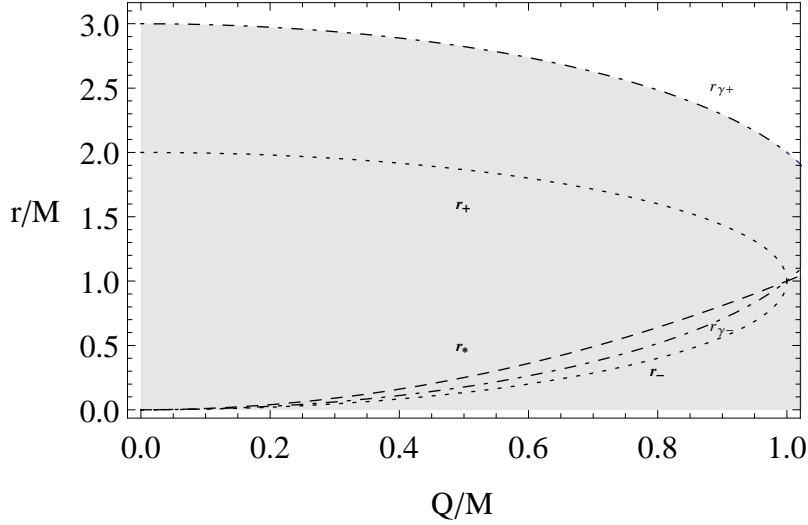}%&
%\includegraphics[scale=.6]{PcEqstab1.eps}\\(a)&(b)\\
%\includegraphics[scale=.6]{PcEqstab1m.eps}\\(c)
%\end{minipage}Plotphi0bpcritico
\end{tabular}
\caption{In this graphic the radii
$r_{\gamma_{+}}\equiv[3M+\sqrt{(9M^2-8Q^2)}]/2$ and $r_{\pm}=
M+\sqrt{M^2-Q^2}$, and $r_* = Q^2/M$ are plotted.
Timelike circular orbits exist only
for $r>r_{\gamma_{+}}$, whereas $r=r_{\gamma_{+}}$ represents
a null hypersurface. Circular motion inside the regions $r<r_*$ and $r\in
(r_{\gamma_{-}},r_{\gamma_{+}})$ is possible only
along spacelike geodesics. }
\label{danielabh}
\end{figure}

From the expressions for the energy and angular momentum of a
timelike particle $(k=1)$ we see that motion is possible only for
$r>Q^2/M=r_*$ and for $r^2-3Mr+2Q^2>0$, i. e., $r<r_{\gamma_{-}}$ and
$r>r_{\gamma_{+}}$, with
$r_{\gamma_{\pm}}\equiv[3M\pm\sqrt{(9M^2-8Q^2)}]/2$. In fact, from
Eqs.(\ref{23}) and (\ref{df}) it follows that the motion inside the
regions $r<r_*$ and $r\in (r_{\gamma_{-}},r_{\gamma_{+}})$ is
possible only along spacelike geodesics. At $r=r_{\gamma_{+}}$ one
finds instead that the velocity of test particles, as defined in
Appendix%\.
\ref{AppA}, is $\nu_{g_{+}}=1$, i.e., the circle $r=r_{\gamma_{+}}$
represents a null hypersurface. On the other hand, for a
non-vanishing charge in the black hole region one can show that
\be
r_- < r_{\gamma_{-}}<r_*< r_+< r_{\gamma_{+}},
 \quad\mbox{with }\quad r_{+}=r_{-}= r_{\gamma_{-}}=r_*\quad\mbox{for}\quad Q=M
\ee
\be
\quad\mbox{and}\quad r_{-}= r_{\gamma_{-}}=r_*
\quad\mbox{for}\quad Q=0 \ .
\ee
The location of these radii for
different values of the mass-to-charge ratio is depicted in
Fig.\il\ref{danielabh}. In the black hole region we limit
ourselves to the study of circular timelike orbits, i.e., orbits with $r>r_{\gamma_{+}}$.

\begin{figure}%[h!]
\centering
\begin{tabular}{cc}
\includegraphics[scale=0.65]{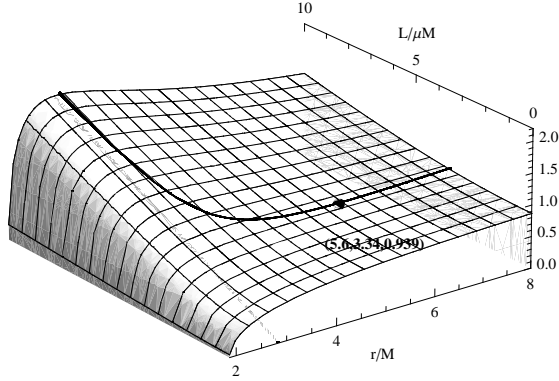}
\end{tabular}
\caption{The effective
potential $V$  for a neutral particle of mass $\mu$ in a
RN  black hole of charge-to-mass ratio $Q/M=0.5$, is
plotted as function of $r/M$  in the range $[1.87,8]$, and the
angular momentum $L/\mu$ in $[0,10]$. In this case the outer horizon
$r_{+}=1.87M$, and $r_{\gamma_{+}}=2.823M$ (see text). Circular orbits
exist for $r>2.823M$. The solid line represents the location of
circular orbits (stable and unstable). The last circular orbit is
represented by a point. The number close to the plotted point represent the
radius $r/M=5.6$, the angular momentum $L/\mu=3.34$ and the energy
$E/M=0.939$  of the last stable circular orbit.}
\label{PlotV0}
\end{figure}

\begin{figure}%[h!]
\centering
\begin{tabular}{cc}
\includegraphics[scale=.6]{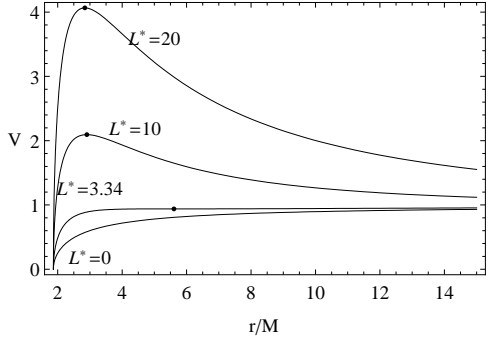}%&
\end{tabular}
\caption{The effective potential $V$ for a neutral particle of mass $\mu$ in a
RN black hole of charge $Q$ and mass $M$ is plotted
in terms of the radius $r/M$  for different values of the angular
momentum $L^{*}\equiv L/(M\mu)$ and $Q=0.5M$. The outer horizon is
located at $r_{+}\approx1.87M$. The effective potential has a
minimum, $V_{min}\approx0.93$, at $r_{min}\approx5.60M$ , for
$L^{*}\approx3.34$.
The plotted points represent local extrema. }
\label{PlotV1}
\end{figure}

The effective potential (\ref{peff}) for a test particle with a fixed value of $Q/M$ is plotted in Fig.\il\ref{PlotV0}, for different values of the angular momentum $L/(M\mu)$.
At infinity, the effective potential tends to a constant  which
is independent of the value of the parameters of the test particle
and of the gravitational source. In our case,
this constant is normalized by choosing the value of the total energy of the
particle as $E/\mu$. Moreover, as the outer horizon is approached
from outside, the effective potential reaches its global minimum
value which is zero. This behavior is illustrated in Fig.\il
\ref{PlotV1} where the effective potential is depicted for a
specific value of $Q/M$ and different values of $L/(M\mu)$.
The radius of a circular orbit, $r_{\ti{CO}}$, is determined by the real positive
root of the equation
\be
 M r^3 -\left(Q^2 +\frac{L^2}{\mu^2}\right) r^2 + \frac{3ML^2}{\mu^2}r - \frac{2Q^2L^2}{\mu^2}=0\ .
\ee
In general, in the region $r>r_{\gamma+}$ circular orbits do not always exist. For instance,
for $Q=0$
circular orbits exist only for values of $|L/(\mu M)|>\sqrt{12}\approx3.45$, whereas for
$Q=M$  and  $Q=0.5M$ the existence condition implies that
$|L/(\mu M)|>\sqrt{8}\approx2.83$ and $|L/(\mu M)|>3.33$, respectively.

\begin{figure}%[h!]
\centering
\begin{tabular}{cc}
\includegraphics[scale=.75]{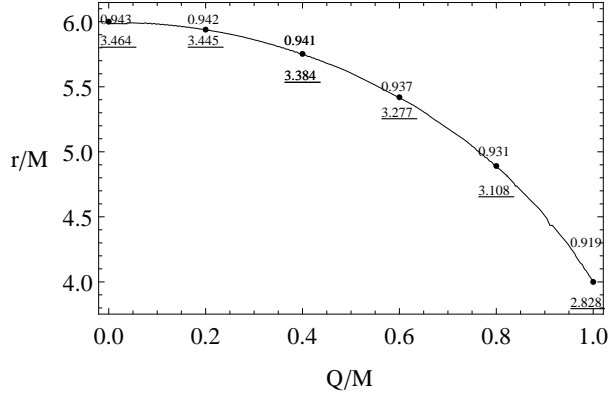}%&
\end{tabular}
\caption{The radius of the last stable orbit $r{\ti{LSCO}}/M$ is plotted as a
function of the ratio $Q/M$.
Numbers close to the points represent the value of the
energy $E/\mu$ and of the angular momentum $L/(\mu M)$ (underlined
numbers) of the corresponding orbit.}
\label{PcEqstab0}
\end{figure}

In this context, it is interesting to explore the stability properties of the circular motion at $r=r_{\ti{CO}}$.
To find the explicit value of the last stable radius we solve the condition (\ref{fgh}) in the black hole region and find
\begin{equation}\label{200920010}
\frac{r_{\ti{LSCO}}}{M}= 2+\frac{4-\frac{3 Q^2}{M^2}+\left[8+\frac{2
Q^4}{M^4}+\frac{Q^2 \left(-9+\sqrt{5-\frac{9 Q^2}{M^2}+\frac{4
Q^4}{M^4}}\right)}{M^2}\right]^{2/3}}{\left[8+\frac{2
Q^4}{M^4}+\frac{Q^2 \left(-9+\sqrt{5-\frac{9 Q^2}{M^2}+\frac{4
Q^4}{M^4}}\right)}{M^2}\right]^{1/3}}\ .
\end{equation}
As expected, in the limiting case $Q\rightarrow 0$, we obtain the Schwarzschild value $r^{\ti{max}}_{\ti{LSCO}}=6M$.
The value of $r_{\ti{LSCO}}$  decreases as $Q/M$ increases, until it reaches
its minimum value  $r^{\ti{min}}_{\ti{LSCO}}=4M$ at $Q/M=1$. The general behavior of $r_{\ti{LSCO}}$ in
terms of the ratio $Q/M$ is illustrated in Fig.\il\ref{PcEqstab0}.

\begin{figure}%[h!]
\centering
\begin{tabular}{cc}
\includegraphics[scale=.5]{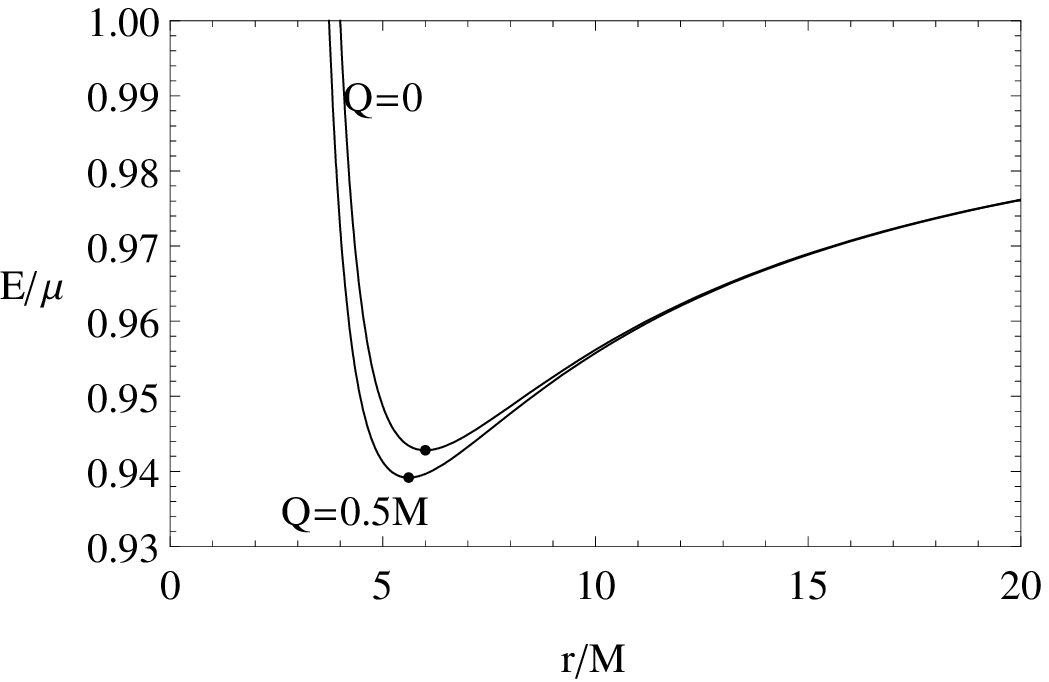}&
\includegraphics[scale=.5]{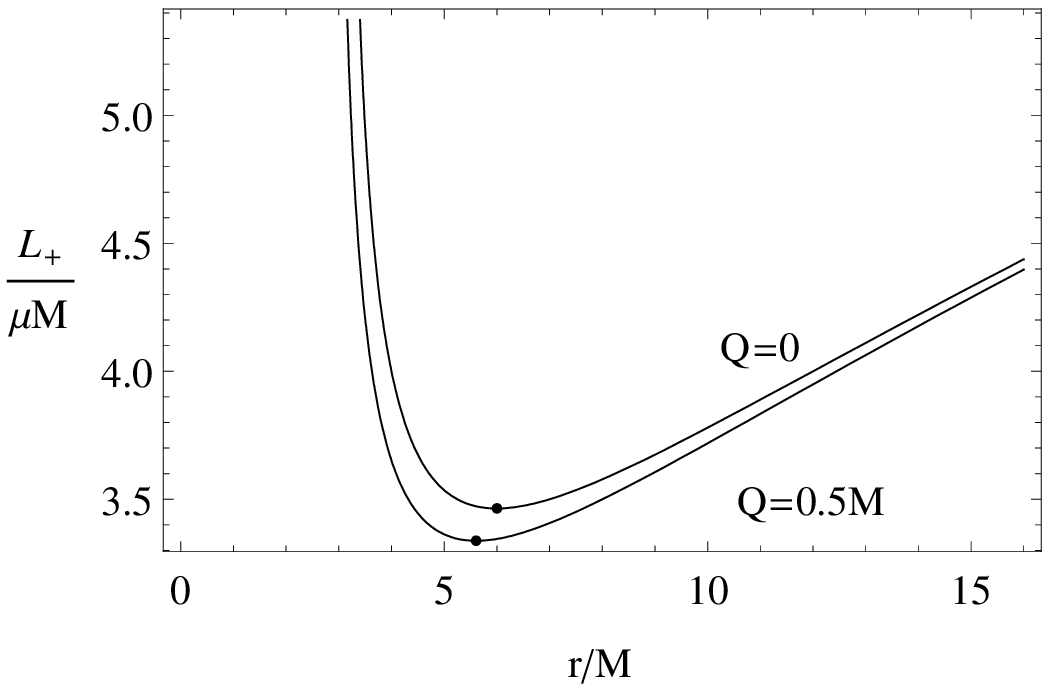}\\(a)&(b)
%\end{minipage}Plotphi0bpcritico
\end{tabular}
\caption{The pictures show
{(a)} the energy $E/\mu$ and {(b)} the angular momentum $L_{+}/(M\mu)\equiv L^*$
of a circular orbit as a function of $r/M$. Notice that for
$Q=0$, the relevant radii are $r_{+}=2M$ and $r_{\gamma_{+}}=3M$, and the minima
are $E_{min}/\mu \approx 0.943$ and $L^{*}_{min}\approx3.46$, with
$r_{min}=6M$. Furthermore, for $Q=0.5M$, we obtain
$r_{+}\approx1.87 M$ and $r_{\gamma_{+}}\approx2.83M$, so that the minima
are $E_{min}/\mu\approx0.939$ and $L^{*}_{min}\approx3.34$, with
$r_{min}\cong5.61M$. The energy and
angular momentum diverge as the limiting radius $r_{\gamma+}$ is approached.}
\label{PlotEplotL}
\end{figure}

\begin{figure}%[h!]
\centering
\begin{tabular}{cc}
%\hline
\includegraphics[scale=0.65]{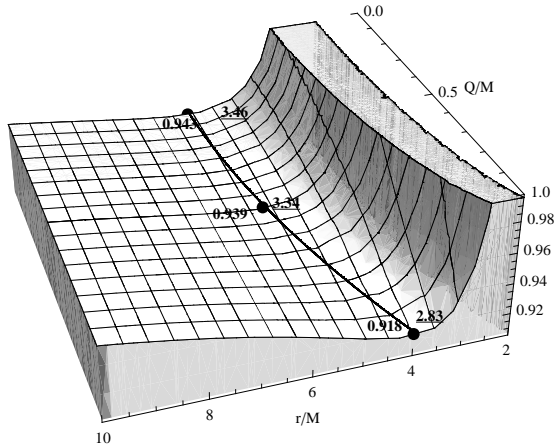}
%\hline
\end{tabular}
\caption{The
energy $E/\mu$ of a test particle on a circular orbit as a function of $r/ M$ and charge-to-mass ratio
$Q/M\in [0,1]$.  The radius of the last stable circular orbit is also plotted (thick
line). Numbers close to the points correspond to  the energy and the
angular momentum (underlined numbers) of the last stable circular
orbit.
} \label{Pa3Q2BH}
\end{figure}

Orbits with $r>r_{\ti{LSCO}}$ are stable. Circular
motion in the region $r_{\gamma_{+}}<r<r_{\ti{LSCO}}$ is completely
unstable. Since the velocity of a test particle at
$r=r_{\gamma_{+}}$ must equal the velocity of light, one can expect
that a particle in the unstable region will reach very rapidly the
orbit at $r=r_{\ti{LSCO}}$. For a static observer inside the
unstable region, the hypersurface $r=r_{\gamma_{+}}$ might appear
 as a source of ``repulsive gravity".
This intuitive result can be corroborated by analyzing the behavior
of energy and angular momentum of test particles. Indeed, Figs.\il\ref{PlotEplotL} shows $E/\mu$ and $L/(M\mu)$ as functions of $r/M$,
for different values of the charge-to-mass ratio of the black hole.
Both quantities diverge as the limiting radius $r=r_{\gamma_{+}}$ is
approached, indicating that an infinite amount of energy and angular
momentum is necessary to reach $r=r_{\gamma_{+}}$.
 As the ratio $Q/M$ increases, the values of the energy and angular momentum at the last
stable orbit decrease. For large values of the radius, the energy of
circular orbits approaches the limit $E=\mu$, and the angular
momentum $L/(M \mu)$ increases monotonically.
A more detailed illustration of this behavior in the case of the energy of the test particle
is represented in Fig.\il\ref{Pa3Q2BH} which shows $E/\mu$ in terms of the ratio $Q/M$ and the
radial distance $r/M$.

\begin{figure}%[h!]
\centering
\begin{tabular}{cc}
\includegraphics[scale=.5]{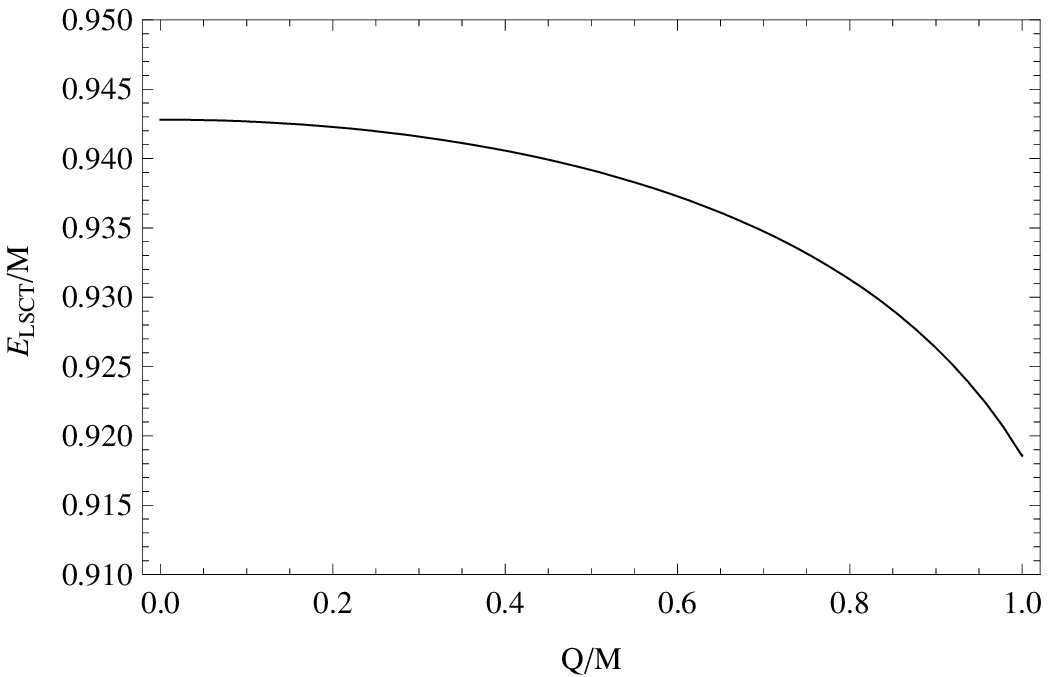}&
\includegraphics[scale=.5]{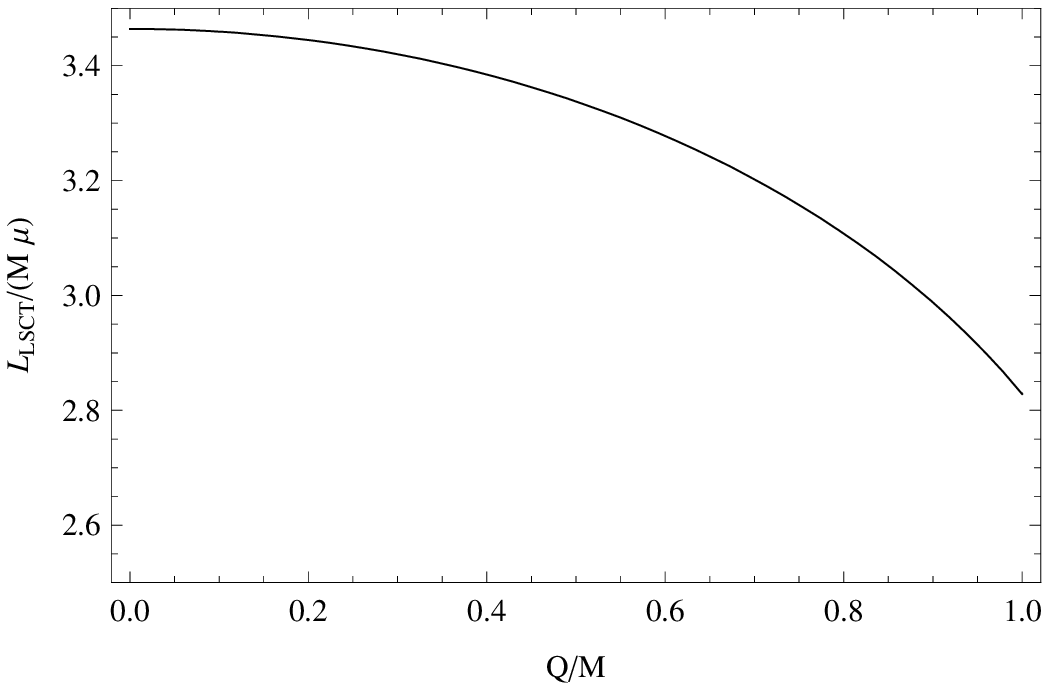}
\\(a)&(b)\\
%\end{minipage}Plotphi0bpcritico
\end{tabular}
\caption[font={footnotesize,it}]{\footnotesize{Plots of the
energy $E_{\ti{LSCO}}/M$ {(a)}, and the angular momentum
$L_{\ti{LSCO}}/(\mu M)$ {(b)} of the last stable circular orbit
in terms of the ratio $Q/M$ of the black hole.} }
\label{PcLastEqstab}
\end{figure}

The analytical expressions for the energy and angular momentum at the last stable orbit
can be obtained by  introducing
the expression (\ref{200920010}) into Eqs.\il(\ref{23}) and (\ref{df}). A numerically analysis of the
resulting expressions show the behavior depicted in Fig.\il(\ref{PcLastEqstab}).

As a general result we obtain that the values for the radius of the last stable orbit as well as of the
corresponding energy and angular momentum diminish due to the presence of the electric charge. Physically, this
means that the additional gravitational field generated by the electric charge acts on neutral particles
as an additional attractive force which reduces the radius of the last stable orbit.

%%%%%%%%%%%%%%%%%%%%%%%%%%%%%%%%%%%%%%%%%%%%%%%%%%%%%%%%%%%%%%
%%%%%%%%%%%%%%%%%%%%%%%%%%%%%%%%%%%%%%%%%%%%%%%%%%%%%%%%%%%%%%
\subsection{Extreme black hole}
\begin{figure}%[h!]
\centering
\begin{tabular}{cc}
\includegraphics[scale=.6]{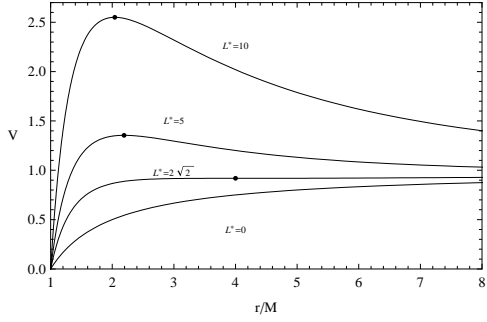}%&
\end{tabular}
\caption[font={footnotesize,it}]{\footnotesize{The effective
potential $V$ for a neutral particle of mass $\mu$ in the field of
an extreme RN black hole  is plotted
as function of the radius $r/M$  for different values of the angular
momentum $L^{*}\equiv L/(M\mu)$. The outer horizon is
located at $r_{+}=M$. For $L^{*}\approx2.83$ the effective potential has a minimum
$V_{min}\approx0.91$ at $r_{min}=4M$. There is
a maximum $V_{max}\approx1.35$ at $r\approx2.19M$ for
$L^{*}=5$, and for $L^{*}=10$ the maximum $V_{max}\approx2.55$ is located at $r \approx2.04M$.
} } \label{PEBH}
\end{figure}

\begin{figure}%[h!]
\centering
\begin{tabular}{cc}
\includegraphics[scale=.55]{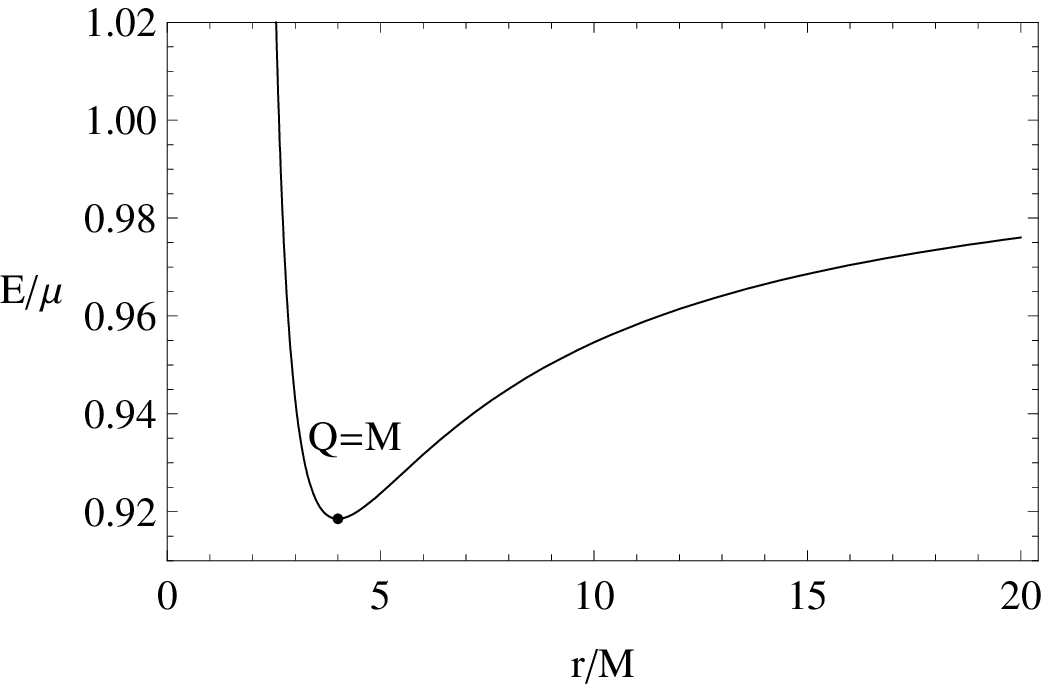}&
\includegraphics[scale=.55]{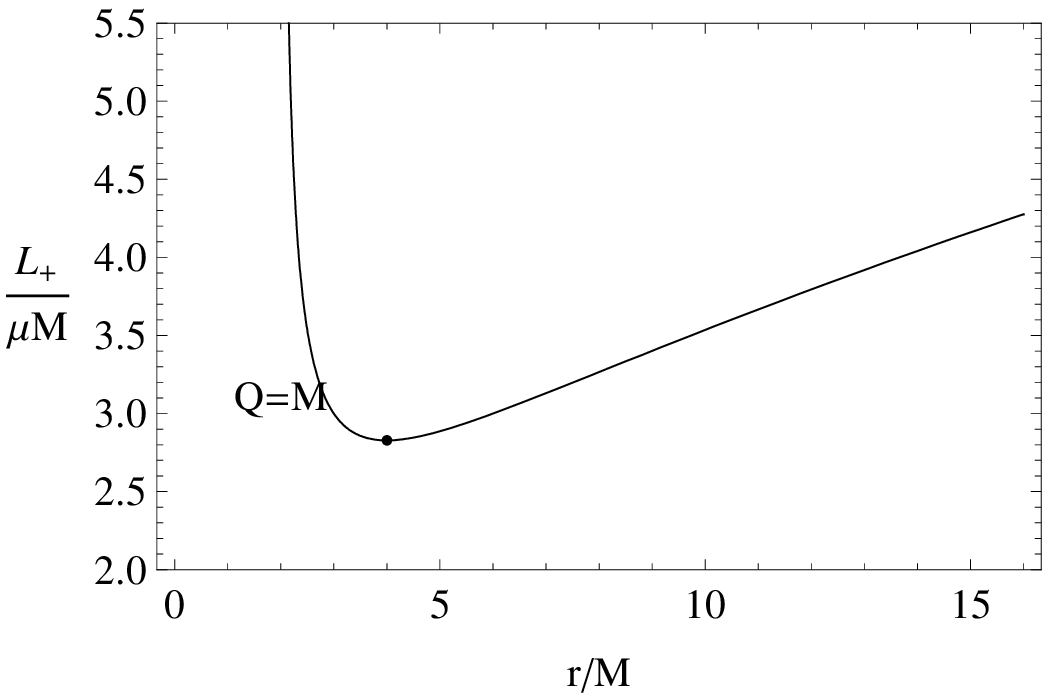}\\(a)&(b)
%\end{minipage}Plotphi0bpcritico
\end{tabular}
\caption[font={footnotesize,it}]{\footnotesize{Plots of {(a)} the energy $E/\mu$ and {(b)}
the angular momentum $L_{+}/(M\mu)\equiv L^{*}$ of a neutral test particle in circular motion around
an extreme RN black hole. The outer horizon
is $r_{+}=M$ and $r_{\gamma_{+}}=2M$}. At $r=4M$ the minimum values of the energy
 $E_{min}/\mu\approx0.91$ and angular momentum $L^{*}_{min}\approx2.83$ are reached. }
\label{PlotEplotLEBH}
\end{figure}

In the case of  an extreme black hole  $(Q=M)$ the outer and inner horizons
coincide at $r_{\pm}=M$. The effective potential vanishes at the horizon and tends to 1
as spatial infinity is approached. In this open interval no divergencies are observed.
This behavior is illustrated in Fig.\il\ref{PEBH} where the effective potential
\be
V=\sqrt{1+\frac{L^2}{\mu^2r^2}} \left(1-\frac{M}{r}\right)
\ee
is plotted for different values of the angular
momentum $ L/(M\mu)$.
%, the effective potential has aminimum for $r_{min}=4M$ $V_{min}\approx0.918$,
For $Q=M$ the radius of circular orbits is
\be \frac{r_{\ti{CO}}}{M}=\frac{L^2-L
\sqrt{-8\mu^2M^2+L^2}}{2\mu^2M^2}\ .
 \ee
The energy and angular momentum of test particles moving along
circular orbits are given by \be \frac{E^2}{\mu^2}=
\frac{(r-M)^3}{r^2(r-2M)} ,\qquad \frac{L^2}{\mu^2}= \frac{M
r^2}{r-2M} \ . \ee Consequently, timelike circular orbits are
restricted by the condition $r>r_ {g+} = 2M$. Figure
\ref{PlotEplotLEBH} shows the behavior of these quantities in terms
of the radial distance. As $r\rightarrow 2M$, the energy and angular
momentum diverge, indicating that the circular motion at
$r_{\gamma_{+}}$ is possible only along null geodesics. As expected,
the local minimum of these graphics determine the radius of the last
stable orbit. At $r=r_{\ti{LSCO}}=4M$, the energy is $E\approx0.918\mu$
and  the angular momentum $L= 2\sqrt{2}\mu M$.

%
%
%%%%%%%%%%%%%%%%%%%%%%%%%%%%%%%%%%%%%%%%%%%%%%%%%
%%%%%%%%%%%%%%%%%%%%%%%%%%%%%%%%%%%%%%%%%%%%%%%%%%
\section{Naked singularity}
\begin{figure}%[h!]
\centering
\begin{tabular}{cc}
\includegraphics[scale=0.6]{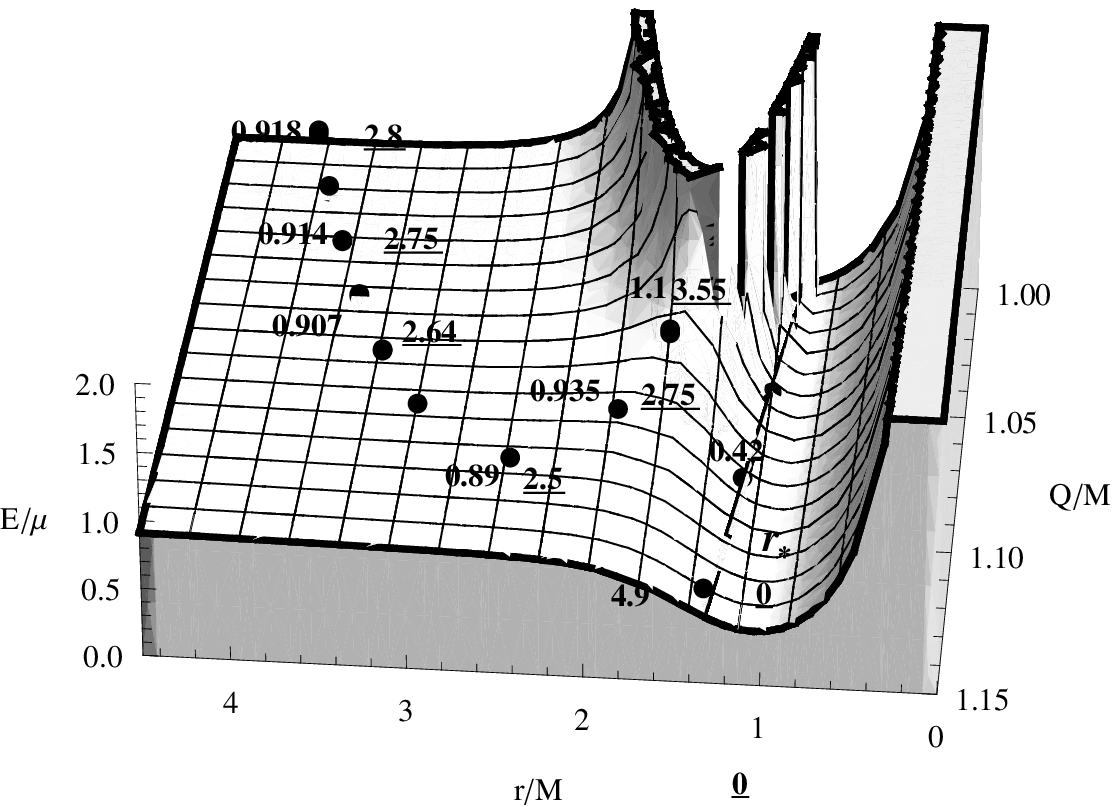}&
\includegraphics[scale=0.6]{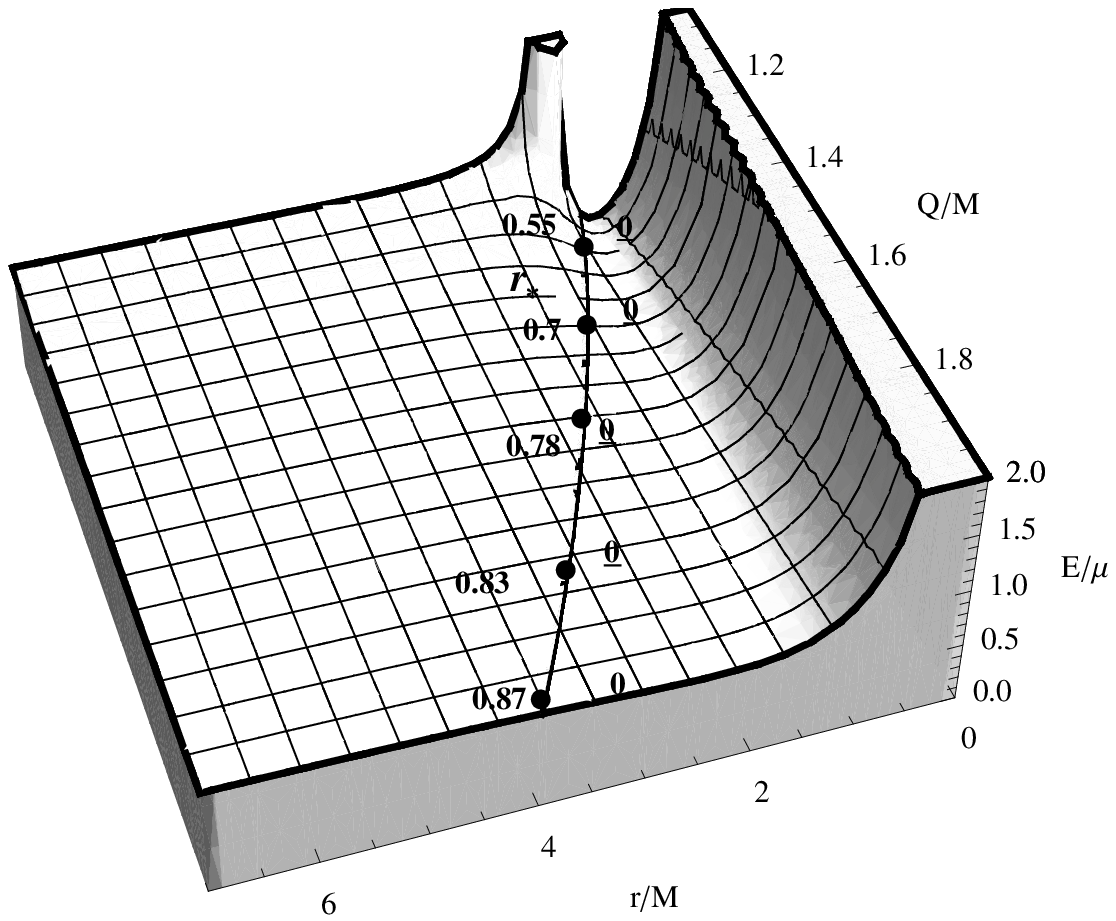}
\\(a)&(b)\\
%\hline
\end{tabular}
\caption[font={footnotesize,it}]{\footnotesize{Plot of the
energy $E/\mu$ of a neutral particle in circular motion around a RN naked singularity
as a function of $r/ M$ and $Q/M$ in the  interval  $[1,1.6]$ {(a)} and in the interval $[1.1,2]$
{(b)}. The line $r_{*}=  Q^2/M$  is also plotted. Numbers
close to the points represent the energy and the angular momentum
(underlined numbers) of the last stable circular orbits.}
}\label{Pa3Q2ns}
\end{figure}

In the naked singularity case $Q>M$ and the energy and angular momentum of the test particle can be
written as
\begin{equation}
\label{St200NS}
\frac{E}{\mu}=\frac{r^2 -2Mr + Q^2}{r \sqrt{r^2-3Mr + 2Q^2 }},
\quad\frac{L_{+}}{\mu }=+ {r} \sqrt{\frac{Mr-Q^2}{r^2-3Mr+2Q^2 }}\ .
\end{equation}
In Fig.\il\ref{Pa3Q2ns} a three-dimensional plot shows the energy as a function
of both the circular orbits radius and the charge-to-mass ratio of the
black hole.

\begin{figure}%[h!]
\centering
\begin{tabular}{cc}
\includegraphics[scale=.7]{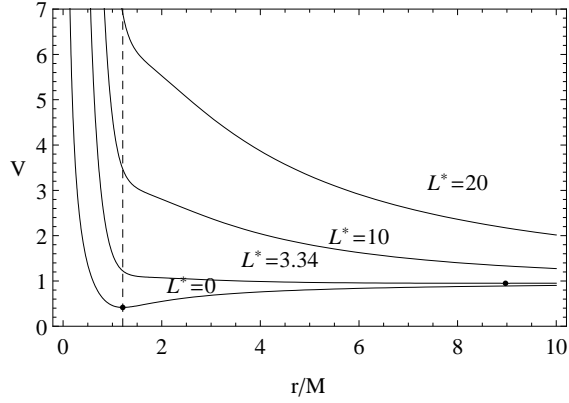}
\end{tabular}
\caption[font={footnotesize,it}]{\footnotesize{The effective
potential $V$ for a neutral particle of mass $\mu$ in a
RN naked singularity with $Q/M=1.1$ is plotted
as a function of the radius $r/M$  for different values of the angular
momentum $L^{*}\equiv L/(M\mu)$.
The classical radius $r_{*}\equiv Q^2/M=1.21M$ is represented by a dashed line.
For $L^{*}=0$ the effective potential presents a minimum $V_{min}\approx0.42$
at $r_{min}=r_{*}$. For $L^{*}\approx3.37$ the minimum $V_{min}/\mu\approx0.95$
is situated at $r_{min}\approx8.97M$.}}
\label{PlotVns11}
\end{figure}

\begin{figure}%[h!]
\centering
\begin{tabular}{cc}
\includegraphics[scale=.7]{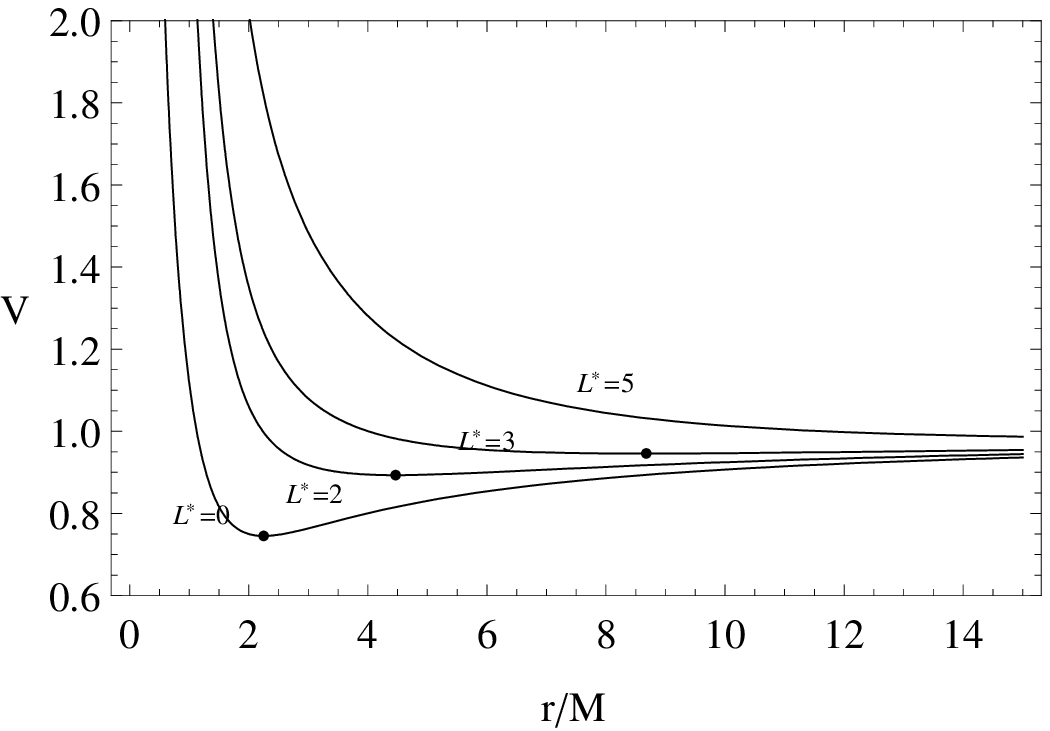}&
\includegraphics[scale=.7]{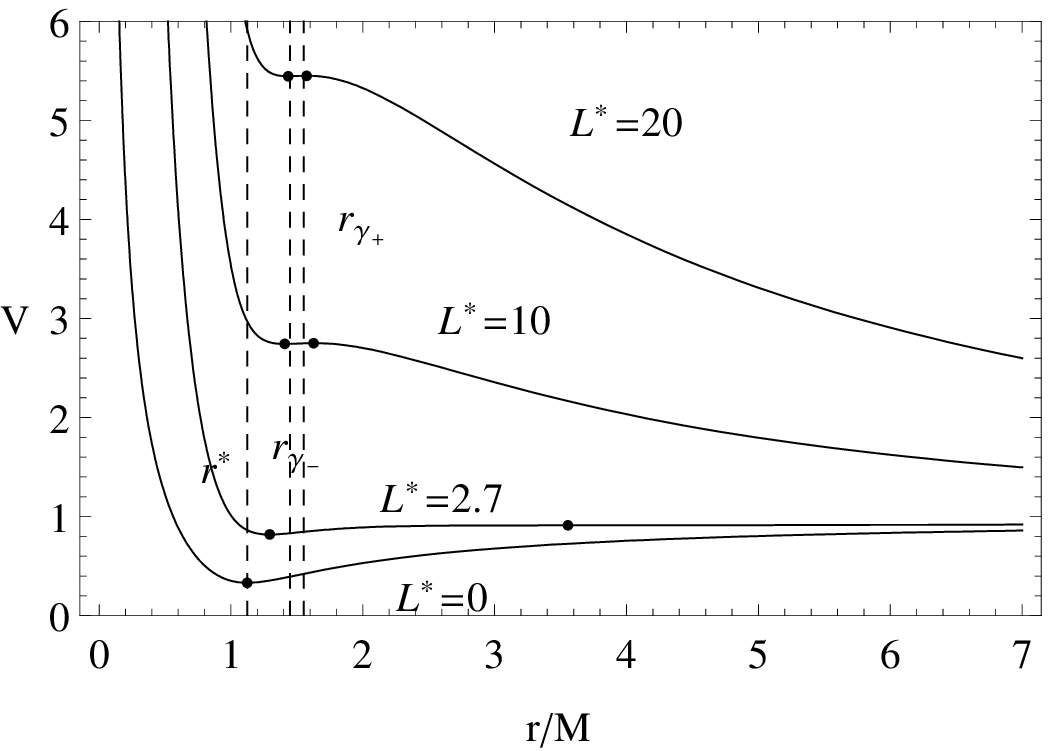}\\(a)&(b)\\
\end{tabular}
\caption[font={footnotesize,it}]{\footnotesize{The effective
potential $V$ for a neutral particle of mass $\mu$ in a
RN naked singularity.
Plot (a) is for $Q/M=1.5$ so that $r_*=Q^2/M = 2.25M.$ For the lowest value
$L^*=0$ at $r=r_*$ there is global minimum with $V_{min}\approx 0.74$. The value
of $V_{min}$ increases as $L^*$ increases. Plot (b) corresponds to $Q/M= 1.06$, and shows
the characteristic radii $r_*=1.12M$, $r_{\gamma_{+}}\equiv[3M+\sqrt{(9M^2-8Q^2)}]/2\approx5.55 M$ and
$r_{\gamma_{-}}\equiv[3M-\sqrt{(9M^2-8Q^2)}]/2\approx1.44 M$. For $L^*=0$ the global minimum is at $r=r_*$
with $V_{min}\approx 0.33$. As $L^*$ increases, the value of $V_{min}$ increases, and at $L^*\approx 2.7$ a
second local minimum appears. The first one with $V_{min}\approx 0.81$ is at $r_{min}\approx 1.29M$ and the
second one with $V_{min}\approx0.91$ is located at $r_{min}\approx3.5M$.
}}
\label{PlotV2}
\end{figure}

From the expressions for energy and angular momentum we see that it is necessary
to consider four different cases: the value $r=r_*=Q^2/M$,
the region inside the interval $1<Q^2/M^2<9/8$, the value
 $Q^2/M^2=9/8$, and finally the region defined by $Q^2/M^2>9/8$.
In Figs.\il\ref{PlotVns11}-\ref{PlotV2} the behavior of the effective potential is exemplified
for different values of the ratio $Q/M$.

%{8.970233674724318`, 0.9496793351779071`}
%

Notice that for a RN naked singularity ($Q/M>1$) the following inequality holds
\begin{equation}
r_{*}\leq r_{\gamma_{-}}\leq r_{\gamma_{+}}\ ,
\end{equation}
where $r_{\gamma_{\pm}}\equiv[3M\pm\sqrt{(9M^2-8Q^2)}]/2$ are the radii at which the value of the
angular momentum and the energy of the test particle diverge.

%%%%%%%%%%%%%%%%%%%%%%%%%%%%%%%%%%%%%%%%%%%%%%%%%%%%%%%%%%%%%%%%%%%%%%%%%
%%%%%%%%%%%%%%%%%%%%%%%%%%%%%%%%%%%%%%%%%%%%%%%%%%%%%%%%%%%%%%%%%%%%%%%%%
\subsection{Static test particles}
\label{sec:sta}

Consider the orbit at
$r=r_{*}$. In the naked singularity case the time-like condition for the
velocity $\nu_g$ is satisfied only for $r\geq r_{*}$ (see Eq.
(\ref{DP})). In the limiting case $r=r_{*}$, a time-like orbit with
\begin{equation}
L(r_{*})=0\ , \quad \frac{E}{\mu}=\sqrt{1-\frac{M^2}{Q^2}} \
\end{equation}
is allowed. It is interesting to note that $r_*=Q^2/M$ coincides with the value of the classical
radius of an electric charge which is usually obtained by using a completely different
approach. This value appears here as the radius at which a particle
can remain ``static" with respect to an observer at infinity. This is an interesting situation
which can be explained intuitively only by assuming the existence of a ``repulsive'' force. The above expression
for the energy of the particle indicates that only in the case of a naked singularity a real value
for the energy can be obtained. In fact, in the case of a black hole, the radius $r_*$ is situated inside
the outer horizon so that $r_*$ cannot reached by classical test particles.
We conclude that  the ``repulsive" force can be the dominant gravitational force only
in the case of a naked singularity.

%%%%%%%%%%%%%%%%%%%%%%%%%%%%%%%%%%%%%%%%%%%%%%%%%%%%%%%%%%%
%%%%%%%%%%%%%%%%%%%%%%%%%%%%%%%%%%%%%%%%%%%%%%%%%%%%%%%%%%%
\subsection{The interval $1<Q^2/M^2<9/8$}
\label{sec:first}

\begin{figure}%[h!]
\centering
\begin{tabular}{cc}
\includegraphics[scale=.55]{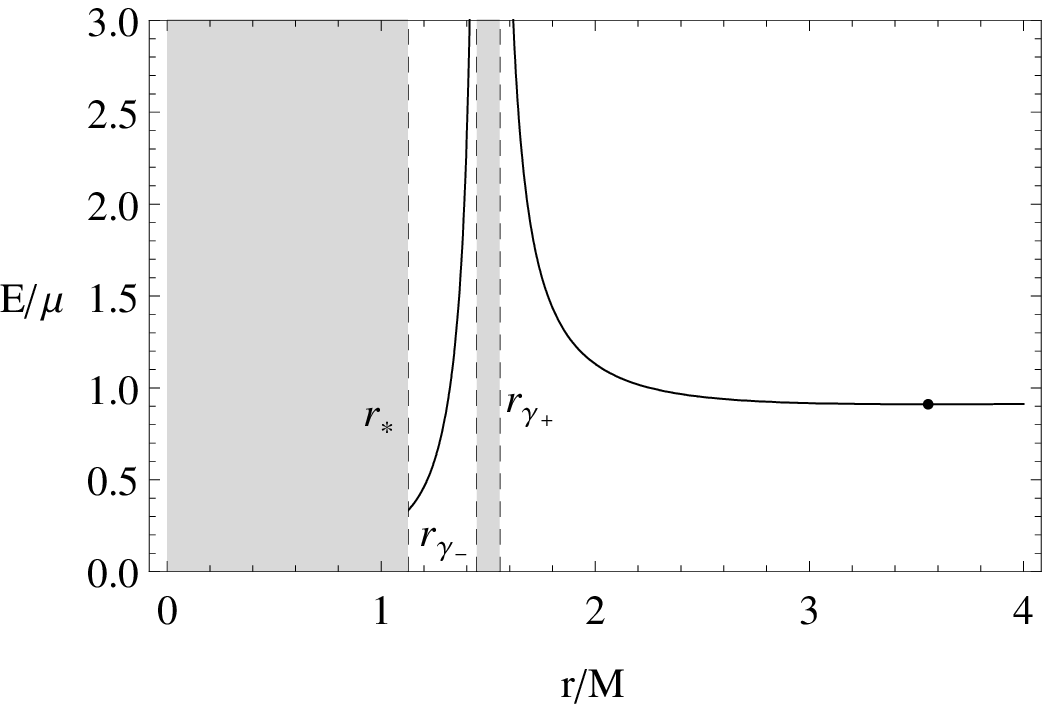}&
\includegraphics[scale=.55]{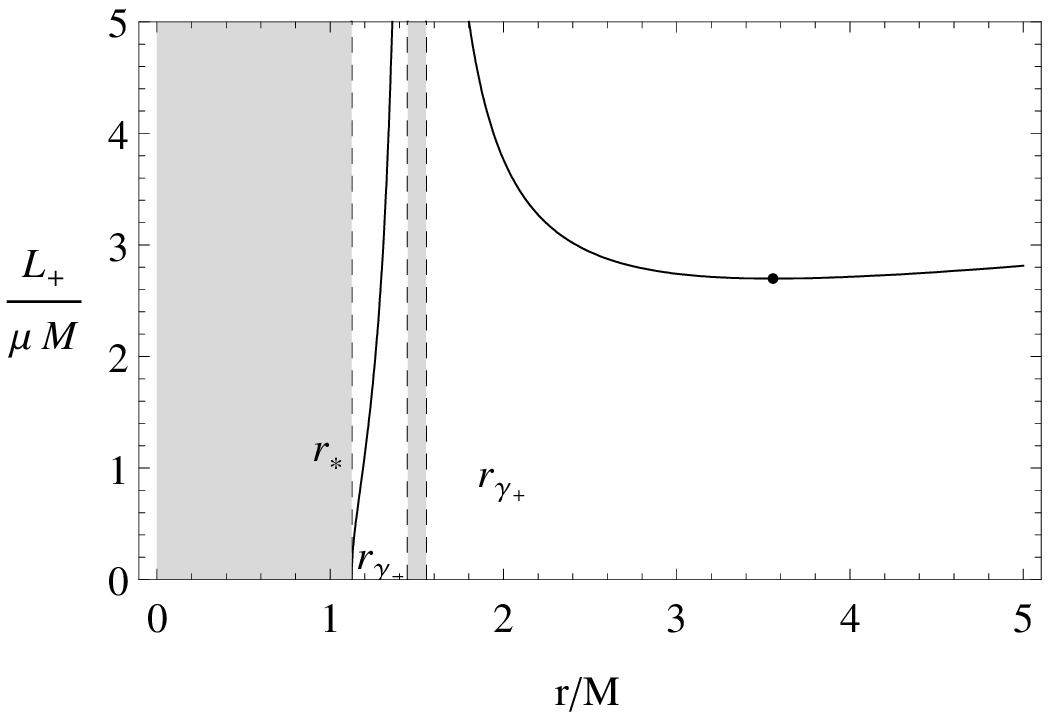}
\end{tabular}
\caption[font={footnotesize,it}]{\footnotesize{The pictures show
{(a)} the energy $E/\mu$ and {(b)} the angular momentum
$L_{+}/(M\mu)\equiv L^*$ of a neutral particle moving on circular orbit around a
RN naked singularity with $Q/M=1.06<\sqrt{9/8}$.
The shaded regions
are forbidden, i.e., circular orbits can exist only in the regions defined by
$r_{*}<r<r_{\gamma_{-}}$ and $r>r_{\gamma_{+}}$, where $r_*=
Q^2/M\approx1.12M$, $r_{\gamma_{-}}\equiv[3M-\sqrt{(9M^2-8Q^2)}]/2\approx1.44M$, and
$r_{\gamma_{+}}\equiv[3M+\sqrt{(9M^2-8Q^2)}]/2\approx1.55M$ (see
text). A minimum for the energy and the angular momentum is located
at $r_{min}\approx3.55M$, where $L^*_{min}\approx2.69$ and
$E_{min}/\mu\approx0.91$. Moreover,  $L^{*}(r_{*})=0$ and
$(E_{min}/\mu)(r_{*})\approx0.33$.} }
\label{PlotEns2}
\end{figure}

In the first region, for $1<Q^2/M^2<9/8$ (see Fig.\il\ref{PlotV2}b),
time-like circular orbits can exist in the
regions $r_{*}< r< r_{\gamma_{-}}$ and $r>r_{\gamma_{+}}$. The boundaries
$r= r_{\gamma_{\pm}}$ correspond to null geodesics, as can be seen from
the expression for the velocity along circular orbits as defined in
Appendix\il\ref{AppA} ($\nu_{g_{\pm}}=1)$. This implies that there are
two regions defined by $r<r_*$ and $r \in [r_{\gamma_{-}},r_{\gamma_{+}}]$
where no time-like particles can be found. This behavior is schematically
illustrated in Fig.\il\ref{PlotEns2}. In the limiting case $Q^2/M^2\rightarrow 1$, the classical
radius coincides with $r_{\gamma-}$ and therefore the only  particle that can
remain ``static" on the classical radius must be a photon.

%%
%%%%%%%%%%%%%%%%%%%%%%%%%%%%%%%%%%%%%%%%%%%%%%%%%%%%%%%%%%%
%%%%%%%%%%%%%%%%%%%%%%%%%%%%%%%%%%%%%%%%%%%%%%%%%%%%%%%%%%%
\subsection{The case $Q^2/M^2=9/8$}
\label{sec:sec}

\begin{figure}%[h!]
\centering
\begin{tabular}{cc}
\includegraphics[scale=.7]{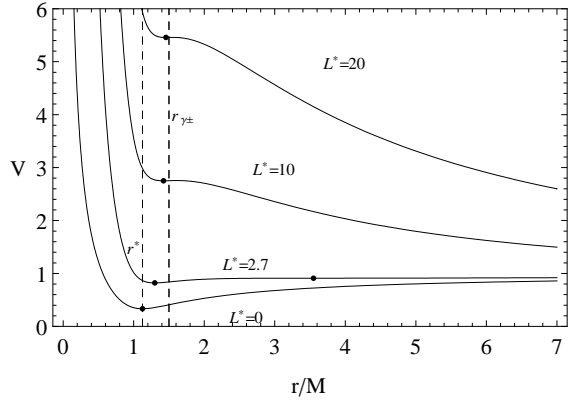}
\end{tabular}
\caption[font={footnotesize,it}]{\footnotesize{The effective
potential $V$ for a neutral particle of mass $\mu$ in a
RN naked singularity with $Q/M=\sqrt{9/8}$  is plotted
as function of the radius $r/M$  for different values of the angular
momentum $L^{*}\equiv L/(M\mu)$. In this case, $r_{*}=  Q^2/M=1.12M$ and
$r_{\gamma_{\pm}}\equiv[3M\pm\sqrt{(9M^2-8Q^2)}]/2 =1.5 M$.
}}
\label{PlotVns98}
\end{figure}

\begin{figure}%[h!]
\centering
\begin{tabular}{cc}
\includegraphics[scale=.7]{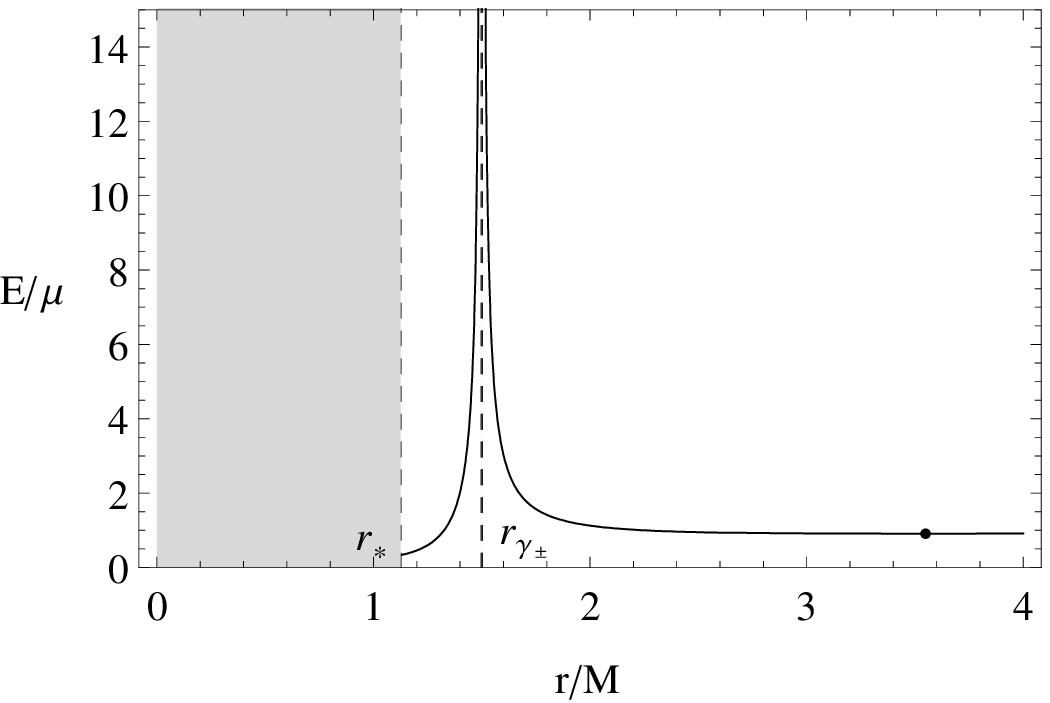}&
\includegraphics[scale=.7]{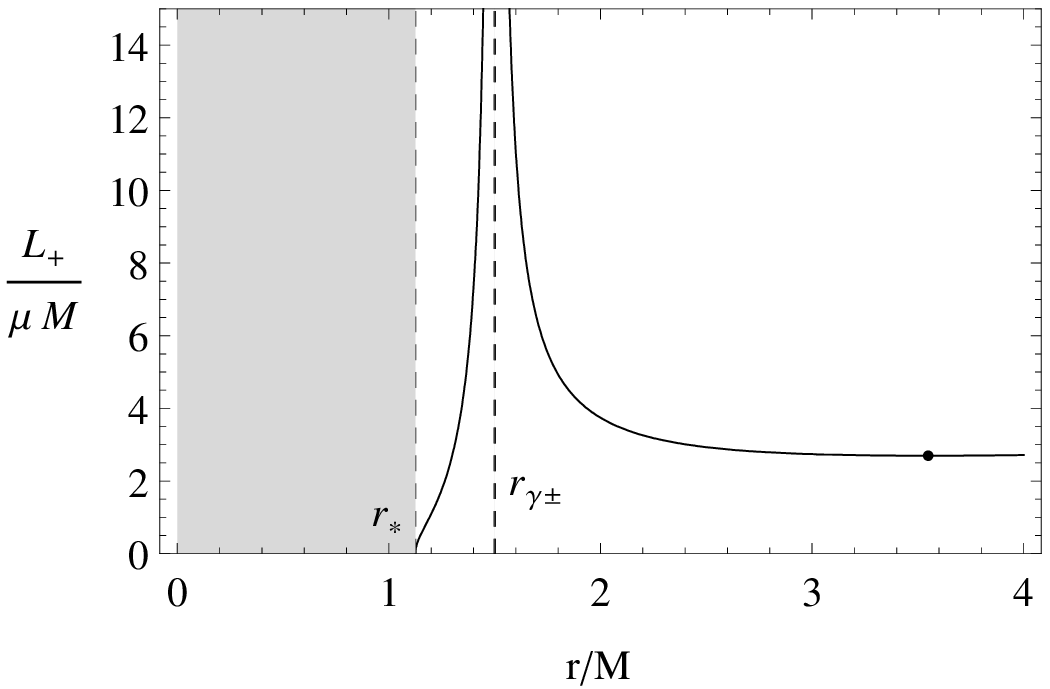}
\end{tabular}
\caption[font={footnotesize,it}]{\footnotesize{Behavior of
{(a)} the energy $E/\mu$ and {(b)} the angular momentum
$L_{+}/(M\mu)\equiv L^*$ of a neutral particle moving along a circular orbit around
a RN naked singularity with $Q/M=\sqrt{9/8}.$
Shaded regions are forbidden for time-like particles.
} } \label{PlotEns298}
\end{figure}

For  $Q^2/M^2=9/8$, the exterior and interior photon orbits situated at $r_{\gamma+}$ and $r_{\gamma-}$ coincide.
The effective potential behaves as illustrated in
Fig.\il\ref{PlotVns98}. Local minima can be found in different regions, depending on the value of the angular momentum
of the test particle. Time-like circular
orbits exists for all $r> r_{*}$, except at $r=r_{\gamma\pm}=3M/2$, which corresponds to
a photon orbit.
The energy and angular momentum of circular orbits are given by
\begin{equation}
\label{dfg098}
\frac{E}{\mu}=\frac{r^2-2Mr+\frac{9}{8}M^2}{r\left(r-\frac{3}{2}M\right)}
,\quad
\frac{L_+}{\mu }=+r\frac{ \sqrt{ M\left(r-\frac{9}{8}M\right) }}
{r-\frac{3}{2}M}\ ,
\end{equation}
and are plotted in Fig.\il\ref{PlotEns298}. As the photon orbits are approached, the particle velocity tends to
$\nu_{g}=1$ and  the energy and angular momentum diverge.
%

%%%%%%%%%%%%%%%%%%%%%%%%%%%%%%%%%%%%%%%%%%%%%%%%%%%%%%%%%%%
%%%%%%%%%%%%%%%%%%%%%%%%%%%%%%%%%%%%%%%%%%%%%%%%%%%%%%%%%%%
\subsection{The region $Q^2/M^2>9/8$}
\label{sec:third}

\begin{figure}%[h!]
\centering
\begin{tabular}{cc}
\includegraphics[scale=.7]{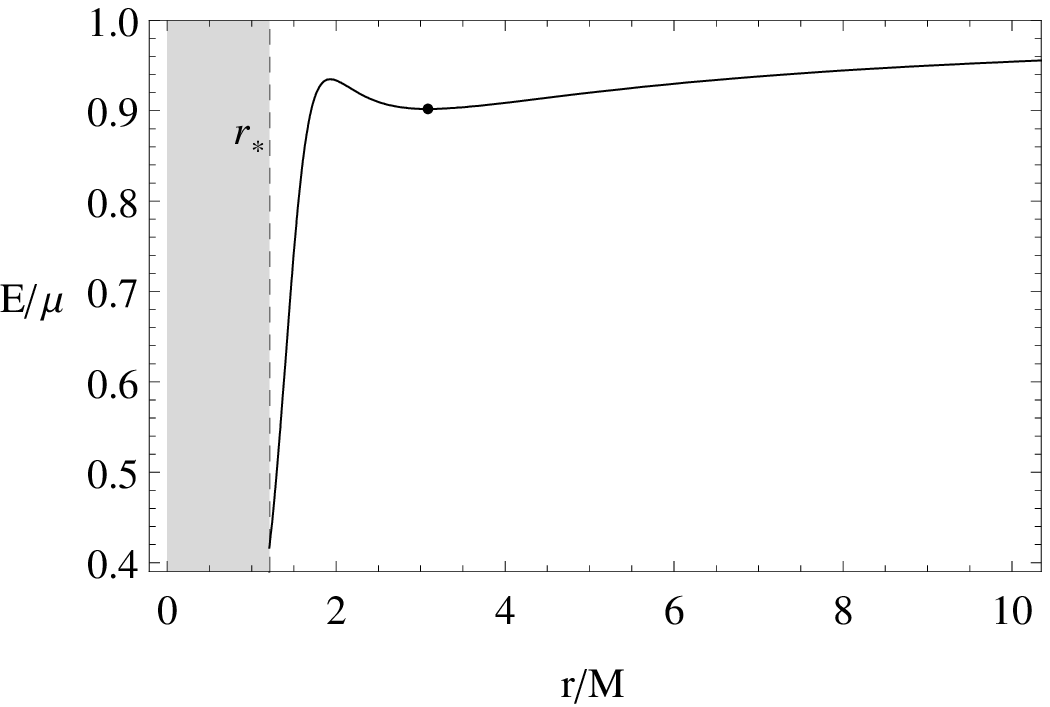}&
\includegraphics[scale=.7]{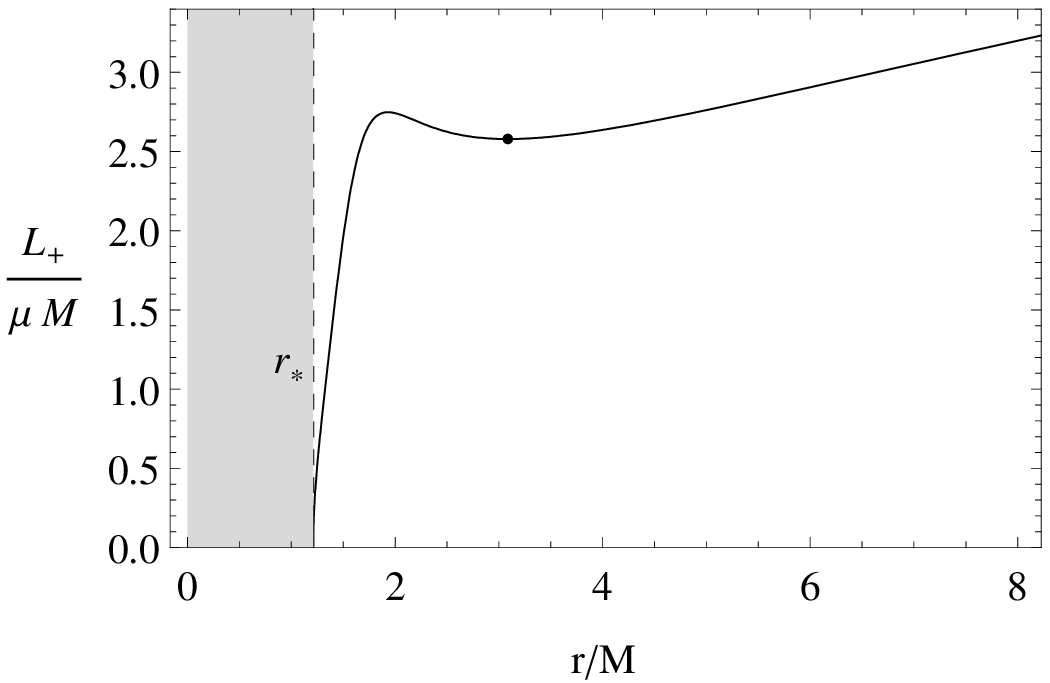}\\\\(a)&(b)\\
\end{tabular}
\caption[font={footnotesize,it}]{\footnotesize{The pictures show
{(a)} the energy $E/\mu$ and {(b)} the angular momentum
$L_{+}/(M\mu)\equiv L^*$ of a neutral particle moving along a circular orbit in a RN
naked singularity with $Q/M=1.1$ as a function of $r/M$.
The shaded regions are
forbidden. Circular orbits can exist only for $r>r_{*}\approx1.21M$.
A minimum of the energy and the angular momentum is
located at $r_{min}\approx3.08M$, where $L^*_{min}\approx2.57$ and
$E_{min}/\mu\approx0.90$. Moreover, $L^{*}(r_{*})=0$ and
$(E_{min}/\mu)(r_{*})\approx0.33$.} }
\label{PlotEns2b}
\end{figure}

\begin{figure}%[h!]
\centering
\begin{tabular}{cc}
\includegraphics[scale=.7]{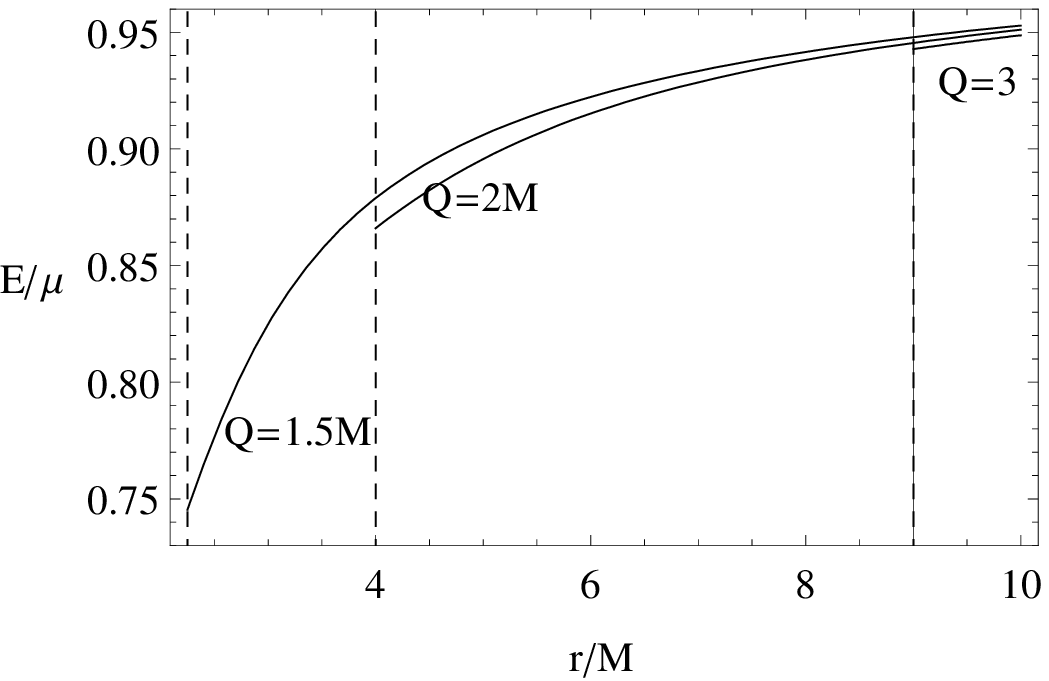}&
\includegraphics[scale=.7]{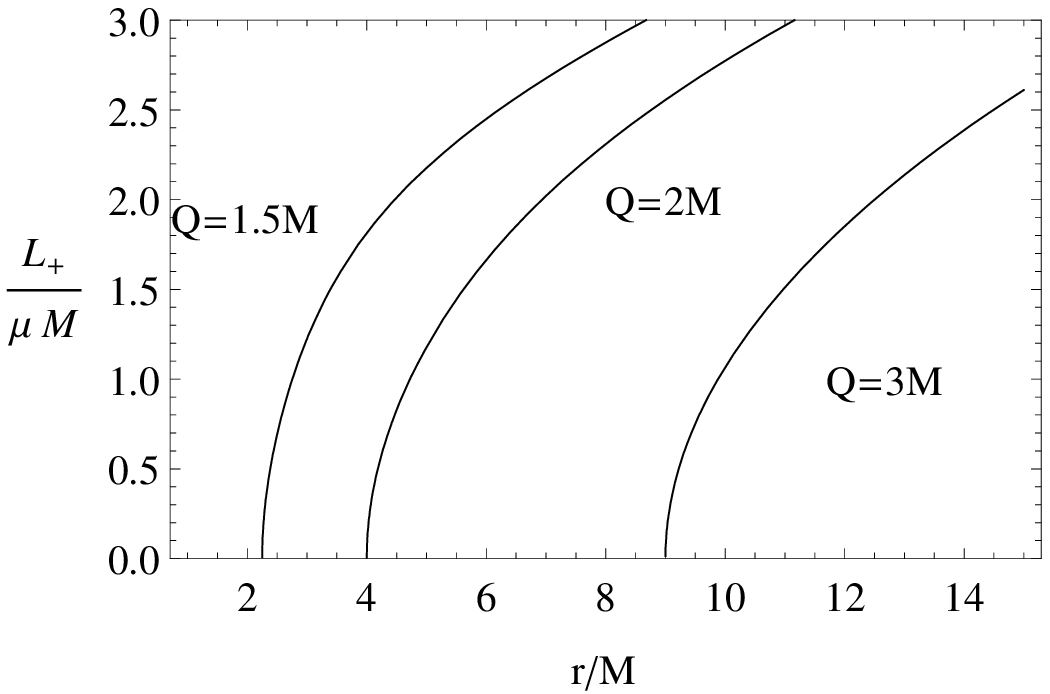}\\(a)&(b)
%\end{minipage}Plotphi0bpcritico
\end{tabular}
\caption[font={footnotesize,it}]{\footnotesize{The pictures show
{(a)} the energy $E/\mu$ and {(b)} the angular momentum
$L_{+}/(M\mu)$ of a test particle of mass $\mu$ in the field of a RN naked singularity
as a function of $r/ M $ and  for
different values of the ratio $Q/M$, satisfying the condition $(Q/M)^2>9/8$ (see text).
Circular orbits can exist only for $r> r_{*}\equiv Q^2/M $.} }
\label{PlotEplotLns}
\end{figure}

For $Q^2/M^2>{9/8}$, the effective potential behaves as illustrated in Figs.\il\ref{PlotVns11} and \ref{PlotV2}a.
Time-like
circular orbits can exist for all $r> r_{*}$. The time-like velocity condition
($\nu_{g}<1$) is always satisfied so that no limiting photon orbits can exist in this case.
The energy and the angular momentum are plotted in
Figs.\il\ref{PlotEns2b} and \ref{PlotEplotLns}.
The angular momentum
increases as the radius of the orbit $r/M$ increases. In the limit of large values of $r$,
the energy tends to $E=\mu$.
%%
%%%%%%%%%%%%%%%%%%%%%%%%%%%%%%%%%%%%%%%%%%%%%%%%%%%%%%%%%%%%%%%%%%%%%%%
%%%%%%%%%%%%%%%%%%%%%%%%%%%%%%%%%%%%%%%%%%%%%%%%%%%%%%%%%%%%%%%%%%%%%%%%
\subsection{The last stable circular orbit}
\label{sec:lsco}

To analyze the stability of circular orbits around a RN naked singularity we solve
Eq.\il(\ref{fgh}) under the assumption that $Q>M$. It turns out that real solutions exist
only in the interval $1<Q/M<\sqrt{5}/2$. They can be represented as
\begin{equation}
\label{sol1}
r^-_{lsco}=2M+2 \sqrt{4M^2-3 Q^2} \cos\left[\frac{1}{3}
\arccos\left(\frac{8M^4-9M^2Q^2+2 Q^4}{M\left(4M^2-3
Q^2\right)^{3/2}}\right)\right]\ ,
\end{equation}
\begin{equation}\label{solx2}
r^+_{lsco}=2M-2 \sqrt{4M^2-3 Q^2} \sin\left[\frac{1}{3}
\arcsin\left(\frac{8M^4-9M^2 Q^2+2 Q^4}{M\left(4M^2-3
Q^2\right)^{3/2}}\right)\right]\ ,
\end{equation}
and
\begin{equation}
\label{solx1}
r_{c}=2M-2\sqrt{4M^2-3 Q^2} \sin\left[\frac{\pi}{6} +\frac{1}{3}
\arccos\left(\frac{8M^4-9M^2 Q^2+2 Q^4}{M\left(4M^2-3
Q^2\right)^{3/2}}\right)\right]\ .
\end{equation}

\begin{figure}%[h!]
\centering
\begin{tabular}{cc}
\includegraphics[scale=.7]{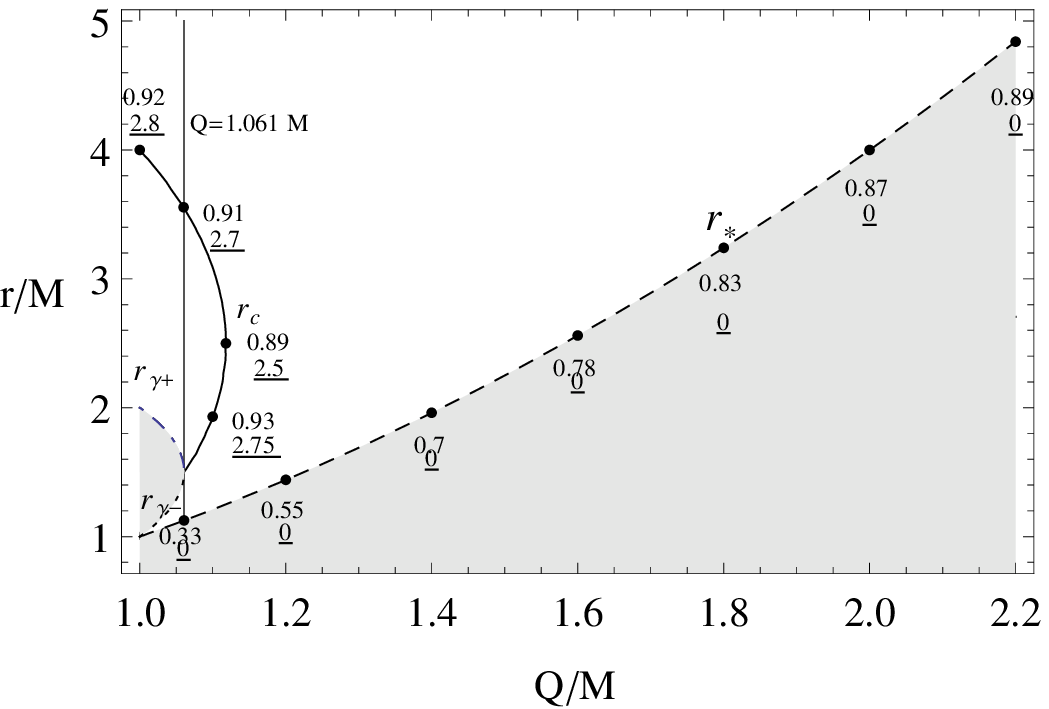}&
\includegraphics[scale=.74]{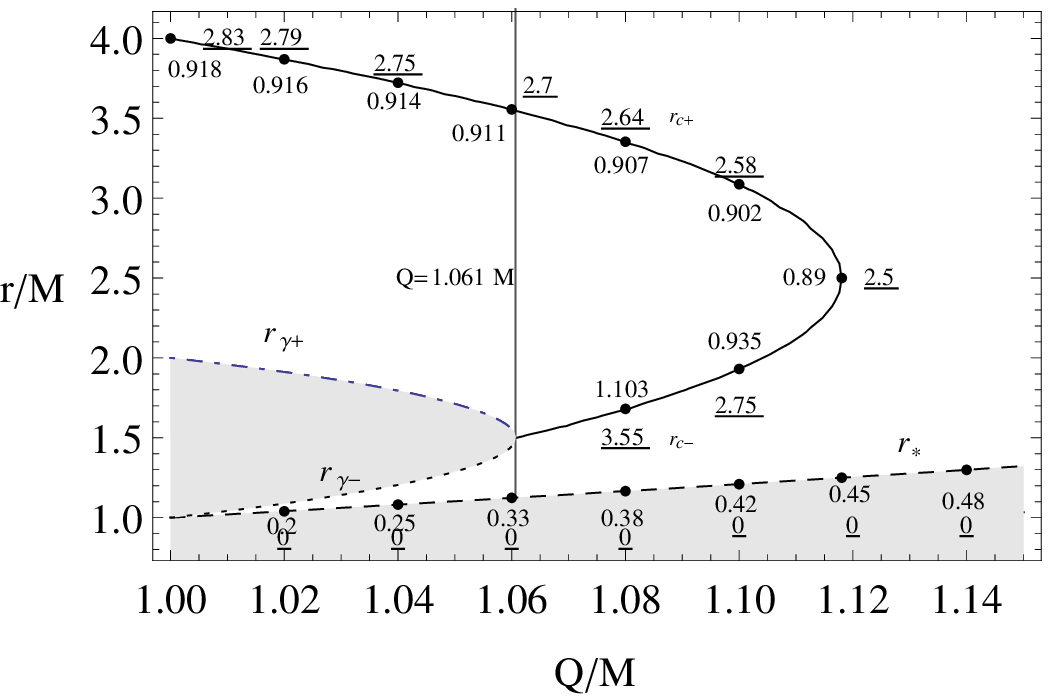}\\(a)&(b)\\
\end{tabular}
\caption{The radius of the last stable circular orbit (solid line)
of a neutral particle moving in the field of a RN naked singularity.
The
ratio $Q/M$ varies in the interval $[1,2.2]$ in plot {(a)}, and  in the interval
$[1,1.15]$ in plot {(b)}. The dashed curve represents the classical radius $r_{*}= Q^2/M$. The dotted
curve corresponds to photon orbits with radius $r_{\gamma_{-}}=[3M-\sqrt{(9M^2-8Q^2)}]/2$,
whereas the dot-dashed curve denotes the photon orbits with radius
$r_{\gamma_{+}}=[3M+\sqrt{(9M^2-8Q^2)}]/2$. Shaded
regions are forbidden.
In the interval $1<Q/M<1.061$ circular orbits can exist
only for $r_{*}<r<r_{\gamma_{-}}$ (all stable) and
$r>r_{\gamma_{+}}$ (unstable for $r_{\gamma_{+}}<r<r^+_{lsco}$ and stable
for $r>r^+_{lsco}$). For $Q/M>1.061$ and $r>r_{*}$ the region of stability
divides into two separated regions: $r_*<r<r^-_{lsco}$ and $r>r^+_{lsco}$.
The numbers close to the plotted
points denote the energy $E/\mu$ and the angular momentum $L/(\mu M)$
(underlined numbers) of the last stable circular orbits.}
\label{PcEqstab1m}
\end{figure}

However, it can be shown that in this interval it holds that
$r_{c}<Q^2/M=r_*$, i. e., this solution is located inside the
classical radius where no time-like circular geodesics are allowed.
Moreover, $r^-_{lsco}<r^+_{lsco}$ in the entire interval, except at $Q/M=\sqrt{5}/2$ where
$r^-_{lsco}=r^+_{lsco}$.
For $Q/M>\sqrt{5}/2$
no solutions of Eq.\il(\ref{fgh}) exist in the region defined by $r>Q^2/M$ so that for
$(Q/M)^2>9/8$ the last stable ``circular'' orbit is located
precisely at $r= r_{*}=Q^2/M$. This  situation  is sketched in Fig.\il\ref{PcEqstab1m}
where it can be seen that the energy  of a particle
located at $r=r_*$ increases as the charge-to-mass ratio increases.
In the interval $1<Q/M<\sqrt{5}/2$, the energy and angular momentum of circular
orbits situated on $r=r^+_{lsco}$ increase as the ratio $Q/M$ increases.
On the contrary, in the interval $\sqrt{9/8}<Q/M<\sqrt{5}/2$,
the energy and angular
momentum of circular orbits with radius $r=r^-_{lsco}$ increase as the ratio
$Q/M$ decreases. This is due to the fact that at $Q/M=\sqrt{9/8}$, the radius $r^-_{lsco}$ coincides
with the radii $r_{\gamma+}$ and $r_{\gamma-}$ which correspond to photon-like orbits.
For
$Q/M>\sqrt{5}/2$, the last stable circular orbit is situated at the classical radius $r_*=Q^2/M$ and the entire
region $r>r_*$ is a region of stability.
For $\sqrt{9/8}<Q/M<\sqrt{5}/2$, there are two regions of
stable orbits, namely, $r_*<r<r^-_{lsco}$ and $r>r^+_{lsco}$, separated by a zone
of instability defined by $r^-_{lsco}<r<r^+_{lsco}$. For $1<Q/M<\sqrt{9/8}$, two
regions of stable orbits appear at $r_*<r<r_{\gamma_{-}}$ and at
$r>r^+_{lsco}$; these regions are separated by the forbidden
region, $r_{\gamma_{-}}<r<r_{\gamma_{+}}$, in which no circular
time-like orbits exist, and by the instability region located at
$r_{\gamma_{+}}<r<r^+_{lsco}$.
%%%%%%%%%%

\begin{figure}%[h!]
\centering
\begin{tabular}{cc}
\includegraphics[scale=.5]{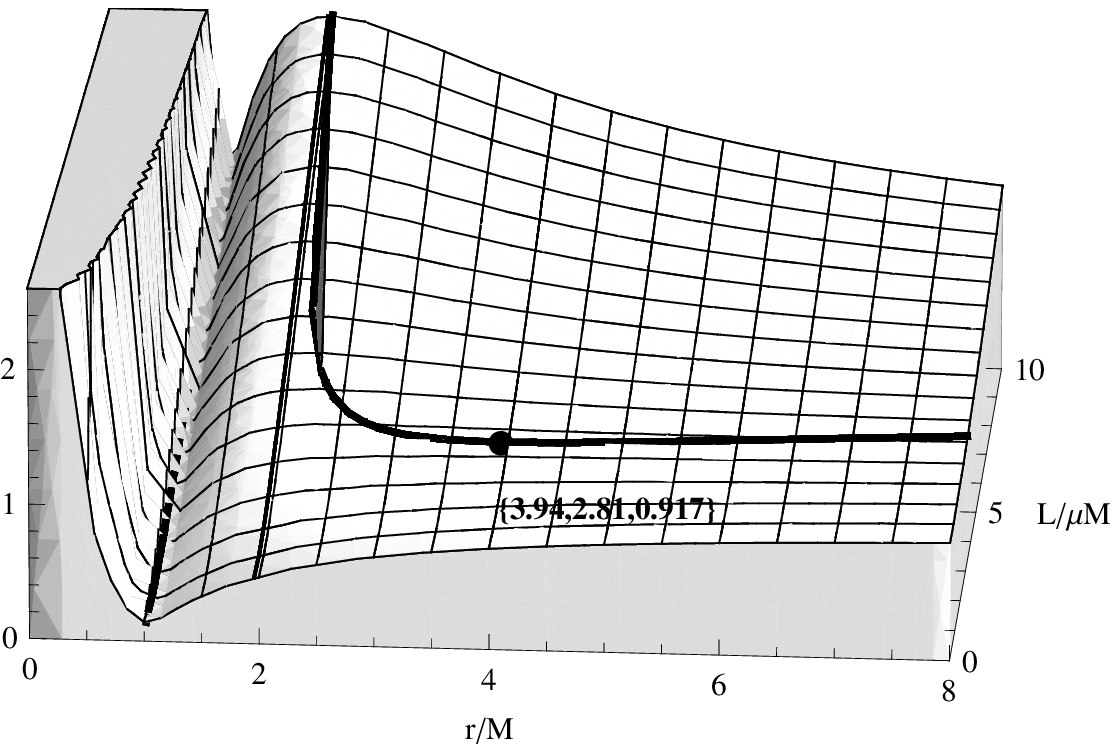}
\includegraphics[scale=.5]{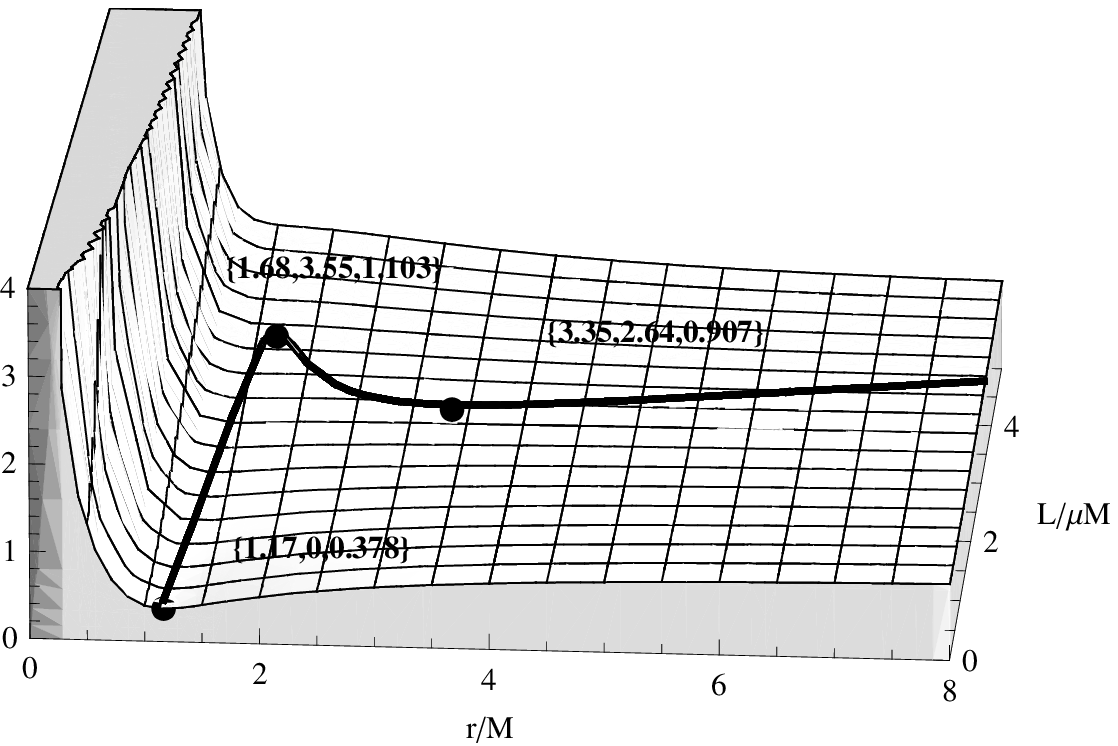}\\(a)\qquad\qquad\qquad(b)\\
\includegraphics[scale=.5]{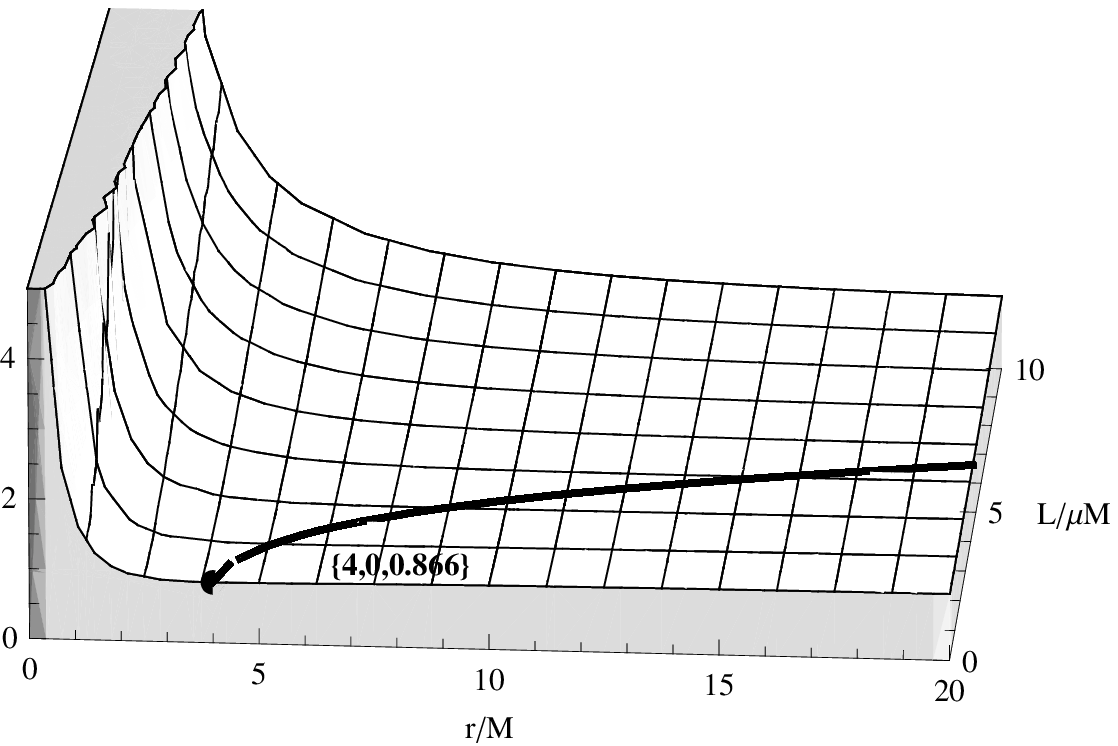}\\(c)\\
\end{tabular}
\caption{The effective
potential $V$ for a neutral particle of mass $\mu$ in a
RN naked singularity is
plotted as a function of the radius $r/M$ and of the  angular momentum
$L^{*}\equiv L/(M\mu)$. Solid lines represent the radii of circular orbits.
The last stable circular orbit radius is denoted by a point. Plot {(a)} is
for $Q=1.1M$ so that $r_{\gamma_{-}}= 1.04M$, $r_{\gamma_+}
=1.95M$. The classical radius is $r_{*}\equiv Q^2/M=1.21M$, where
$L^{*}=0$ and $V_{min}\approx0.42$. Unstable circular
orbits are in $r<3.94M$ and at $r=3.94M$, the values are $L^{*}\approx2.81$
and $V/\mu\approx0.91$. Plot {(b)} is for $Q=1.08M$ so that
$r_{\gamma_{\pm}}$ do not exist and $r_{*}\equiv Q^2/M=1.16M$ with
$V_{min}(r_*)\approx0.37$. Stable orbits are in the interval $r_{*}<r<1.68M$ and at
the boundary, $r=1.68M$, the values are $L^*=3.55$ and $V/\mu=1.10$.
The interval  $1.68M<r<3.35M$ corresponds to unstable orbits with the boundary values $L^*=2.64$ and $V/\mu=0.90$
at $r=3.35M$. Stable orbits are located at $r>3.35M$. Plot {(c)} is for $Q=2M$ and $r_*=4M$. Stable circular orbits
exist for $r>4M$ and at the boundary $V(4M)=0.86\mu$.}
\label{Plot3Vns}
\end{figure}

The curves representing the radii of stable circular orbits   are also showed
in Fig.\il(\ref{Plot3Vns}) where the effective potential is plotted
for selected values of the $Q/M$ as a function of the angular momentum
and the radius.
%
%%%%%%%%%%%%%%%%%%%%%%%%%%%%%%%%%%%%%%%%%%%%%%%%%%%%%%%%%%%%%%%%%%%%%%%%%%%%%%%
%%%%%%%%%%%%%%%%%%%%%%%%%%%%%%%%%%%%%%%%%%%%%%%%%%%%%%%%%%%%%%%%%%%%%%%%%%%%%%%
\section{Black hole versus naked singularity}
\begin{figure}%[h!]
\centering
\begin{tabular}{cc}
%\hline
\includegraphics[scale=0.6]{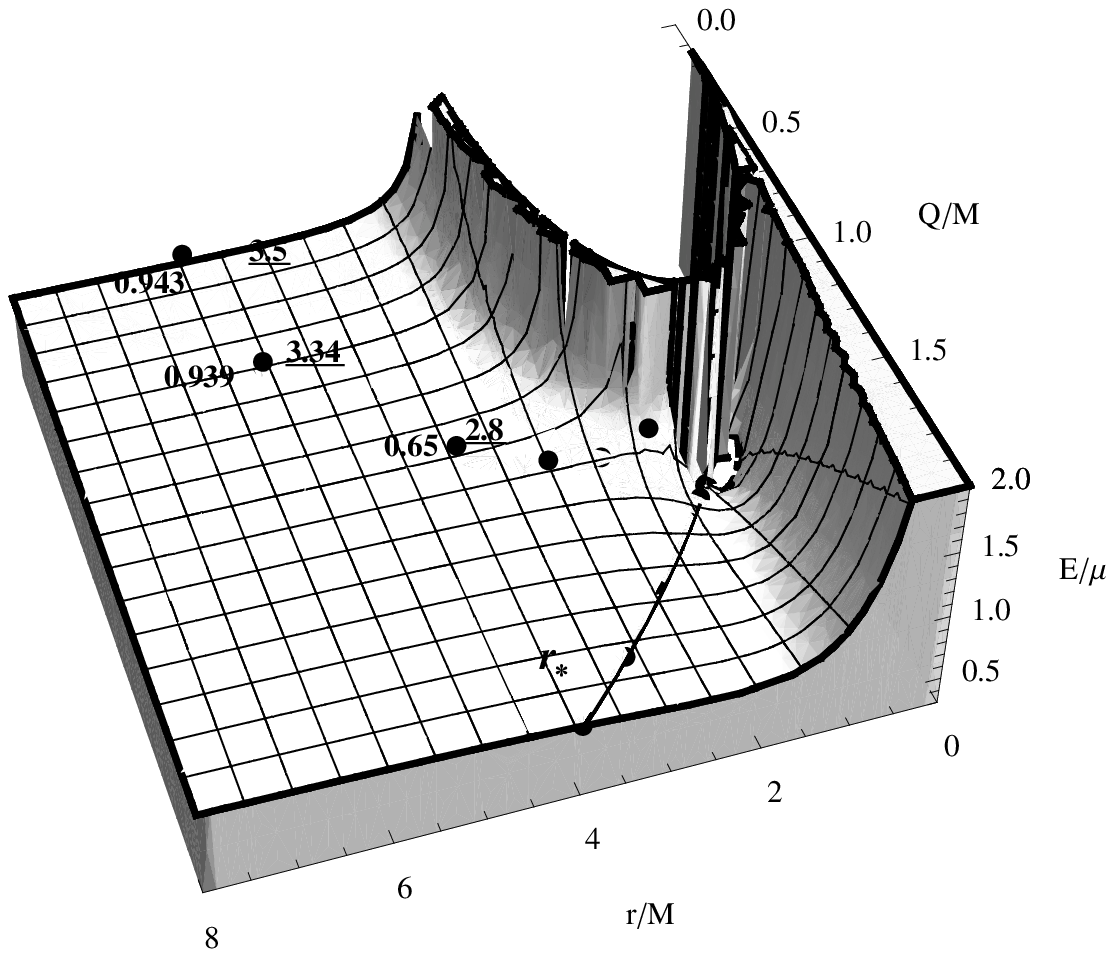}&
\includegraphics[scale=0.6]{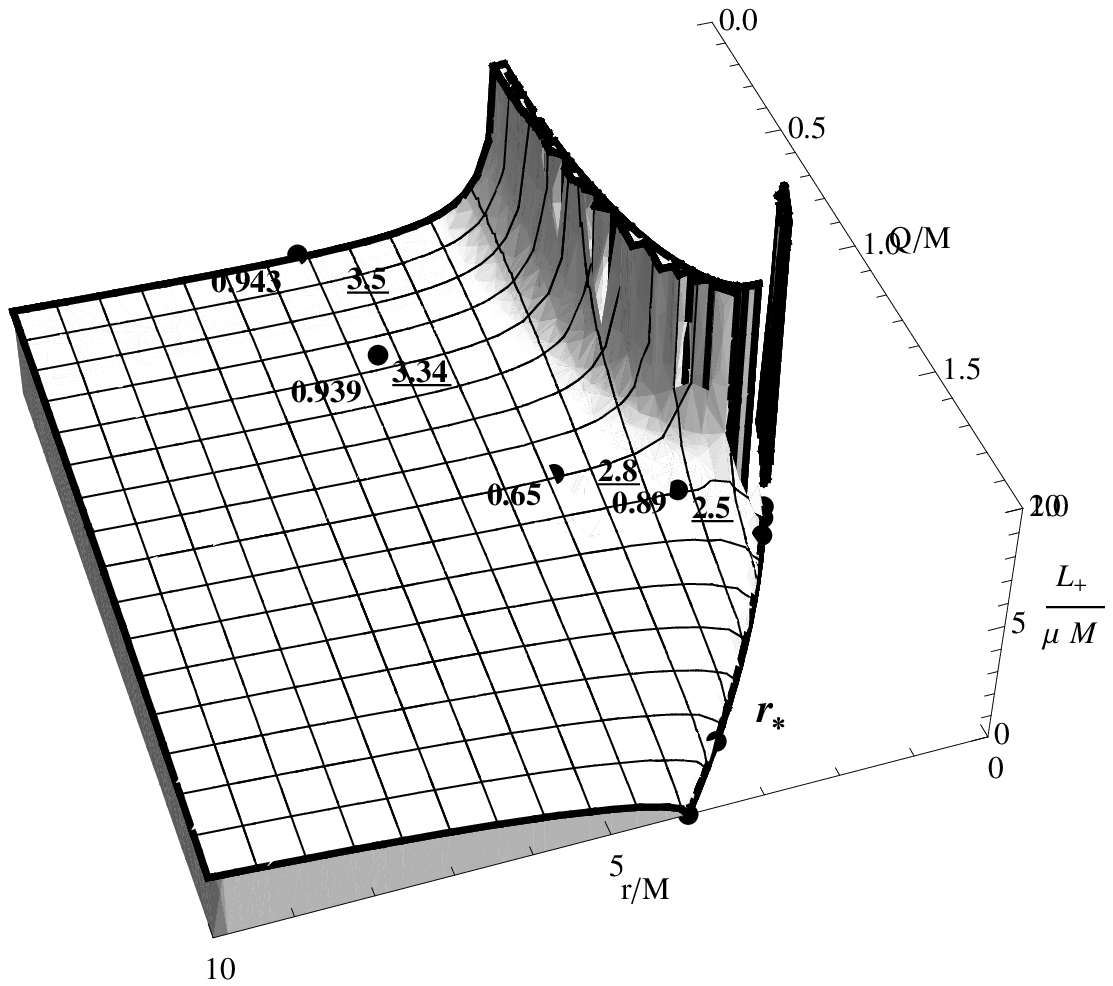} \\(a)&(b)\\
\end{tabular}
\caption{The pictures show the
energy $E/\mu$ {(a)} and the angular momentum $L_{+}/(\mu M)$
{(b)} of a test particle moving along a circular orbit around a RN source as a
function of $r/ M$ and the charge-to-mass ratio $Q/M$ in the  range
$[0,2]$. The curve $r=r_{*}\equiv Q^2/M$ is also plotted. The numbers
close to the points denote the energy and the angular momentum
(underlined numbers) of the last stable circular orbits.}
\label{Pa3Q2tutte}
\end{figure}

\begin{figure}%[h!]
\centering
\begin{tabular}{cc}
%\hline
\includegraphics[scale=1.2]{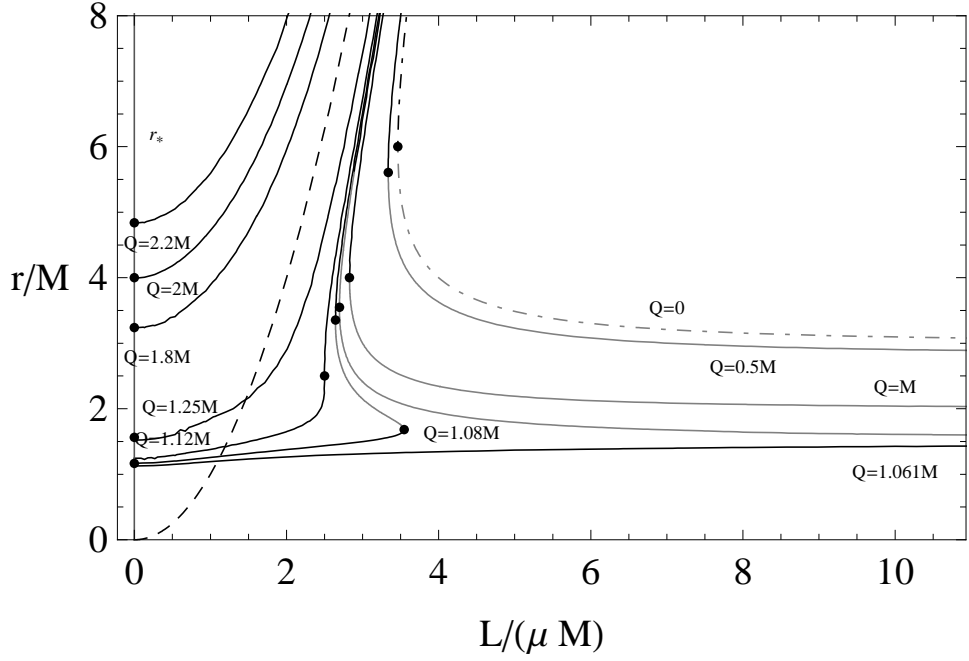}
%\hline
\end{tabular}
\caption{The
radius $r/M$ of circular orbits for a neutral particle of mass $\mu$
in a RN geometry of charge $Q$ and mass $M$ is
plotted as a function of the angular momentum $L/(\mu M)$ for
different values of the charge-to-mass ratio $Q/M$ in the interval $[0,2]$.
The radius of circular orbits for the Newtonian limit (dashed curve) and
for the Schwarzschild spacetime (dot-dashed
curve) are also plotted. The plotted points represent the last stable circular orbits.
Black curves correspond to stable circular orbits, and gray curves denote unstable
circular orbits. The dotted line represents the classical radius $r=r_{*}\equiv Q^2/M$.}
\label{Pcomparativa}
\end{figure}

\begin{figure}%[h!]
\centering
\begin{tabular}{cc}
\includegraphics[scale=1]{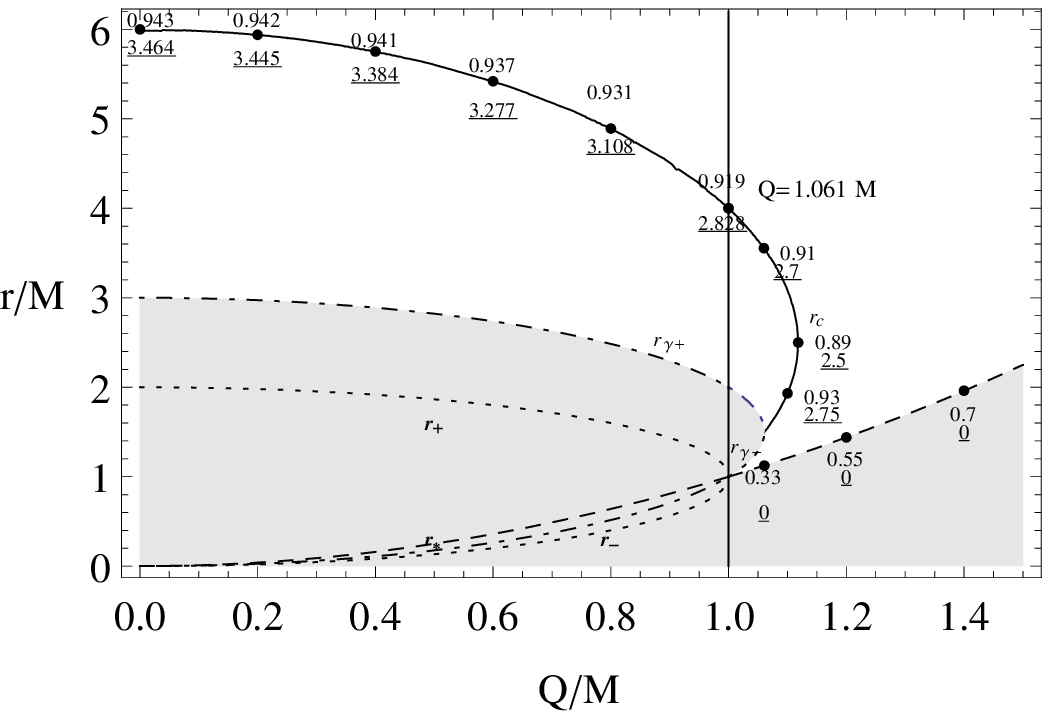}%&
\end{tabular}
\caption{The
 radius $r/M$ of the last stable circular orbit of a neutral
particle of mass $\mu$ in a RN geometry of charge
$Q$ and mass $M$ is plotted as a function of the ratio $Q/M$ in the
interval $[0,1.5]$. Here $r_{\gamma\pm}\equiv[3M\pm\sqrt{(9M^2-8Q^2)}]/2$ and $r_{\pm}\equiv
M\pm\sqrt{M^2-Q^2}$ (see text).
In the Schwarzschild case, $Q=0$,  we find $r_{lsco}=6M$,
$r_{+}= 2M$ and $r_{\gamma_{+}}=3M$. For an extreme black hole, $Q=M$, we find
$r_{lsco}=4M$ with $r_{+}=M$ and $r_{\gamma_{+}}=2M$. The numbers close to the plotted points
denote the
energy $E/\mu$ and the angular momentum $L/(\mu M)$ (underlined
numbers) of the last stable circular orbit.}
\label{PcEqstab}
\end{figure}

The main results obtained from the study of RN black holes and naked
singularities are here summarized and compared in the plots of
Fig.\il\ref{Pa3Q2tutte}--\ref{PcEqstab}.

In the three-dimensional plots  of Fig.\il\ref{Pa3Q2tutte}
the energy and the angular momentum (\ref{df}) are shown as
functions of the circular orbits radius and the  charge-to-mass
ratio $Q/M$, for both black hole and naked singularity cases.

Fig.\il\ref{Pcomparativa} shows the circular orbits radius as a function
of the angular momentum for different black holes and naked singularities.
The cases of a Schwarzschild spacetime and the Newtonian limit are also plotted for
comparison. Clearly, for large values of $r/M$ all the curves converge to the Newtonian limit.
Finally, the study of the circular orbits stability for neutral test
particles in the RN spacetime is summarized in
the Fig.\ref{PcEqstab} where the radius of the  last stable circular
orbits is plotted as function of $Q/M$.
%%%%

The location and structure of the stability regions for neutral
particles around a black hole differ in a very strong way from the case
of a naked singularity.  Black hole sources are characterized by the existence
of only one stability region which can extend from a minimum radius $r_{lsco}\in [4M,6M]$ up to infinity.
In the case of a naked singularity there are two possible scenarios. If the mass-to-charge ratio
satisfies the condition $Q/M\geq \sqrt{5}/2$, there exist only one stability region which extends from
the classical radius $r_*=Q^2/M$ up to infinity. The second scenario appears in naked singularities with
charge-to-mass ratio within the range $1<Q/M<\sqrt{5}/2$. In this case, we have two stability regions
separated by either a region of instability or a forbidden region, where no time-like circular orbits
are allowed, and a region of instability.

%%%%%%%%%%%%%%%%%%%%%%%%%%%%%%%%%%%%%%%%%%%%%%%%%%%%%%%%%%%%%%%%%
%%%%%%%%%%%%%%%%%%%%%%%%%%%%%%%%%%%%%%%%%%%%%%%%%%%%%%%%%%%%%%%%%
\section{Conclusions}
\label{sec:con}
In this work we discussed the motion of neutral test particles along circular
orbits in the RN spacetime. We studied  separately the case of black holes and
naked singularities, emphasizing their differences. Our analysis is based on the
study of the behavior of an effective potential that determines the position and
stability properties of circular orbits. We also investigated in detail the
behavior of the energy and angular momentum of the test particle in all possible
configurations of black holes and naked singularities.

We found that at the classical radius $r=r_*= Q^2/M$ circular orbits exist with ``zero"
angular momentum. This means that a static observer situated at infinity would interpret
this situation as a test particle that remains motionless $(r=Q^2/M=$const, $\phi=$const) as time passes.
This phenomena can take place only in the case of a naked singularity and is interpreted as
a consequence of the ``repulsive" force generated by the charge distribution.

Black holes turn out to be characterized by a single zone of stability which extends from a minimum radius $r_{lsco}$ up to
infinity. The radius of the last stable circular orbit has its maximum value of $r_{lsco}= 6M$ in the Schwarzschild
limiting case, and reaches its minimum value of $r_{lsco}=4M$ in the case of an extreme black hole. The situation is completely
different in the case of naked singularities.
If the mass-to-charge ratio
satisfies the condition $Q/M\geq \sqrt{5}/2$, there exist only one stability region which extends from
the classical radius $r_*=Q^2/M$ up to infinity. In this case, the classical radius determines also the
radius of the last stable orbit that could be, in principle, as large as allowed by the amount of charge $Q$
that can be associated to a given mass $M$. If the
charge-to-mass ratio is within the range $1<Q/M<\sqrt{5}/2$, we found two stability regions that are separated
either by a region of instability, in the interval $ \sqrt{9/8}<Q/M<\sqrt{5}/2$, or a region of instability and
a region forbidden for time-like orbits, in the interval $1<Q/M<\sqrt{9/8}$. This result implies that
around a naked singularity with $1<Q/M<\sqrt{5}/2$ the region of stability splits into two non-connected regions
in which test particles can remain orbiting forever.

%%%%%%%%%%%%%%%%%%%%%%%%%%%%%%%%%%%%%%%%%%%%%%%%%%%%%%%%%%%%%%%%%%%%%%%%%%%%%%%%%%%%
%%%%%%%%%%%%%%%%%%%%%%%%%%%%%%%%%%%%%%%%%%%%%%%%%%%%%%%%%%%%%%%%%%%%%%%%%%%%%%%%%%%%%
\section*{Acknowledgments}
We would like to thank A. Geralico for helpful comments. Two of us (DP and HQ) thank ICRANet support.
This work was supported in part by DGAPA-UNAM, grant No. IN106110.

\appendix
\section{}\label{AppA}
Let us introduce  an orthonormal frame adapted to the static
observers:
\begin{equation}\label{gloria militar}
e_{\hat{t}}=r\Delta^{-1/2}\partial_{t},\quad
e_{\hat{r}}=\frac{\Delta^{1/2}}{r}\partial_{r},\quad
e_{\hat{\theta}}=\frac{1}{r}\partial_{\theta},\quad
e_{\hat{\phi}}=\frac{1}{r\sin\theta}\partial_{\phi}
\end{equation}
with dual
\begin{equation}\label{gloria militar dual}
\omega^{\hat{t}}=\frac{\Delta^{1/2}}{r}dt,\quad
\omega^{\hat{r}}=r\Delta^{-1/2}dr,\quad
\omega^{\hat{\theta}}=rd\theta,\quad \omega^{\hat{\phi}}=r\sin\theta
d\phi
\end{equation}
Consider the  tangent to a (time-like) spatially circular orbit
$u^{a}$ as
$$
u=\Gamma(\partial_t+\zeta\partial_{\phi})=\gamma\left[e_{\hat{t}}+\nu
e_{\hat{\phi}}\right],
$$
where $\Gamma$ and $\gamma$ are two normalization factors:
$$
\Gamma^2=(-g_{tt}-\zeta^2g_{\phi\phi})^{-1}\quad\mbox{and}\quad\gamma^2=(1-\nu^2)^{-1}
$$
which assures that $u_{a}u^{a} = -1$; where $\zeta$ is the angular
velocity with respect to infinity and $\nu$ is the ``local proper
linear velocity" as measured in the frame (\ref{gloria militar}).
The angular velocity $\zeta$ is related to the local proper linear
velocity by
$$
\zeta=\sqrt{-\frac{g_{tt}}{g_{\phi\phi}}}\nu.
$$
so that $\Gamma=\gamma/\sqrt{-g_{tt}}$. Here $\zeta$ and therefore
also $\nu$ are assumed to be constant along the $u$-orbit. As
usually we limit our analysis to the equatorial plane; as a
convention, the physical (orthonormal) component along
$-\partial_{\theta}$, perpendicular to the equatorial plane will be
referred to as along the positive z-axis and will be indicated by
$\hat{z}$ and so $e_{\hat{z}}=-e_{\hat{\theta}}$. Here we focus
attention to the time-like circular geodetic $u_{\pm}$ such that
$\nabla_{u_{\pm}}u_{\pm}=0$, co-rotating ($\zeta_{+}$) and
counter-rotating ($\zeta_{-}$) with respect to the assumed positive
(counter-clockwise) variation of the $\phi$-angle respectively. It
results:
\begin{equation}\label{RR}
\zeta_{\pm}\equiv\pm\zeta_{g}=\pm\frac{\sqrt{(Mr-Q^2)}}{r^2}
\end{equation}
so that
\begin{eqnarray}\label{DP}
\nu_{g}&=&\sqrt{\frac{Mr-Q^2}{\Delta}},\quad\gamma_{g}=\left(\frac{\Delta}{r^2-3Mr+2Q^2}\right)^{1/2}
\\\nonumber\\\label{nobel!}
u_{\pm}&=&\frac{\gamma_{g}}{\sqrt{-g_{tt}}}(\partial_t\pm\zeta_{g}\partial_{\phi})=\gamma_{g}\left[e_{\hat{t}}\pm
\nu_{g} e_{\hat{\phi}}\right],
\end{eqnarray}
with time-like condition $\nu_{g}<1$ satisfied.

The particle's four--momentum is given by $p^{a} = \mu u^{a}$ where
$u^{a}$ is given by (\ref{nobel!}). The conserved quantities
associated with temporal and azimuthal Killing vectors $\xi_{t}$ and
$\xi_{\phi}$ are respectively
$$
p_{\alpha}\xi^{\alpha}_{t}=-\mu\sqrt{-g_{tt}}\gamma_{g}=-E; \quad
p_{\alpha} \xi^{\alpha}_{\phi}=\pm \mu \frac{\gamma_{g}
\zeta_{g}r^{2}}{\sqrt{-g_{tt}}}=L_{\pm}
$$
where E and L are the particle's energy and angular momentum
respectively. Substituting the (\ref{DP}) we infer:
\begin{equation}\label{EQ}
\frac{E}{\mu}=\frac{\Delta}{r\sqrt{r^2-3Mr+2Q^2}},
\quad\frac{L_{\pm}}{M\mu} =\pm
\frac{r\sqrt{r-\frac{Q^2}{M}}}{\sqrt{M}\sqrt{r^2-3Mr+2Q^2}}
\end{equation}
these results are in agreement with the results obtained by the
study of the effective potential.
 
%%%%%%%%%%%%%%%%%%%%%%%%%%%%%%%%%%%%%%%%%%%%%%%%%%%%%%%%%%%%%%%%%%%%%%%%
%%%%%%%%%%%%%%%%%%%%%%%%%%%%%%%%%%%%%%%%%%%%%%%%%%%%%%%%%%%%%%%%%%%%%%%%
%%%%%%%%%%%%%%%%%%%%%%%%%%%%%%%%%%%%%%%%%%%%%%%%%%%%%%%%%%%%%%%%%%%%%%%%
\end{document}